\documentclass[]{aa}

\usepackage{graphicx}
\usepackage{txfonts}
\usepackage{savesymbol}	
\savesymbol{negmedspace}
\savesymbol{negthickspace}
\savesymbol{iint}
\savesymbol{iiint}
\savesymbol{iiiint}
\savesymbol{idotsint}
\usepackage{graphicx}	
\usepackage{amsmath}	
\usepackage{amssymb}	
\usepackage{xcolor}
\usepackage{fixmath}
\usepackage{wasysym}
\usepackage{oplotsymbl}
\usepackage{bm}
\usepackage{xfrac}
\savesymbol{float}

\begin{document}

\title{The impact of 
disk-locking on convective turnover times of low-mass pre-main sequence 
and main sequence stars}
\author{
N.\,R.\ Landin\inst{1,2}
\and
L.\,T.\,S.\ Mendes\inst{3,2}
\and
L.\,P.\,R.\ Vaz\inst{2}
\and S.\,H.\,P.\ Alencar\inst{2}
}
\offprints{N.\,R.\ Landin}
\institute{Universidade Federal de Vi\c{c}osa, Campus UFV Florestal, CEP 35690-000 -- Florestal, MG, Brazil\\
\email{nlandin@ufv.br}
\and
Depto.\ de F\'{\i}sica, Universidade Federal de Minas Gerais, C.P.702, 31270-901 -- Belo Horizonte, MG, Brazil
\and
Depto.\ de Engenharia Eletr\^onica, Universidade Federal de Minas Gerais, C.P.702, 31270-901 -- Belo Horizonte, MG, Brazil
}
\date{Received ; accepted }


\abstract
{}
{The impact of disk-locking on the stellar properties related to magnetic activity 
from the theoretical point of view is investigated.}
{We use the {\ttfamily ATON} stellar evolution code to calculate 
theoretical values of convective turnover times ($\tau_{\rm c}$) and Rossby numbers ($Ro$,  
the ratio between rotation periods and $\tau_{\rm c}$) for pre-main sequence (pre-MS) and main 
sequence (MS) stars.
We investigate how $\tau_{\rm c}$ varies with the initial rotation period and 
with the disk lifetime, using angular momentum conserving models and models simulating 
the disk-locking mechanism. In the latter case, the angular velocity is kept constant, during a 
given locking time, to mimic the magnetic locking effects of a circumstellar disk.}
{The local convective 
turnover times generated with disk-locking models are shorter than those obtained with 
angular momentum conserving models. 
The differences are smaller in the early pre-MS, increase with stellar age and become more accentuated
for stars with $M$\,$\geq$$1\,{\rm M}_{\odot}$ and ages greater than 100\,Myr.
Our new values of $\tau_{\rm c}$ were used to estimate $Ro$ for a sample of stars 
selected from the literature in order to investigate 
the rotation-activity relationship. 
We fit the data with a two-part power-law function and find the best fitting parameters of 
this relation.}
{The differences we found between both sets of models suggest that 
the star's disk-locking phase properties affect its Rossby number and its position in the 
rotation-activity diagram.
Our results indicate that the dynamo efficiency is lower
for stars that had undergone longer disk-locking phases.}


\authorrunning{Landin et al.}
\titlerunning{Convective turnover times of disk-locking  stars}
   \maketitle

\section{Introduction}

The convective turnover time is a typical timescale for convective motions
of the stars' conductive plasma, responsible for transportation of 
a fraction of the energy in some types of stars, like the low-mass ones. When different 
parts of a magnetic star rotate differentially, the interaction between 
convective motions and differential rotation produces the dynamo effect 
\citep[as proposed by][]{parker55} that keeps and regenerates stellar magnetic fields. 
In solar type stars, such fields are thought 
to be created and amplified at the tachocline, a thin shear layer between the 
radiative core and the convective envelope, first defined by \citet{spiegel92}. This mechanism is 
supposed to drive stellar magnetic activity, expressed as a variety of 
observable phenomena like coronal heating, star spots, activity cycles, 
flares and chromospheric and coronal emissions.
For solar-like stars, it is believed that stellar magnetism and rotation 
are regulated by a dynamo process called the $\alpha$$-$${\mathrm \Omega}$ dynamo 
\citep{mohanty03}, in which the poloidal and toroidal field components 
sustain themselves through a cyclic feedback process \citep{nelson08}.
Nowadays, the connection between magnetic activity and rotation is well 
established, and \citet{skumanich72} was the first to suggest that the 
rotation-activity relation is a consequence of the dynamo action and 
showed that the stellar angular velocity, $\mathrm \Omega$, decreases with age as $t^{-1/2}$. 
The rotation-magnetic activity relation has been widely used to investigate
stellar magnetism. It is expressed by the relation between the stellar 
rotation rate (the projected rotational velocity in the line of sight or the rotation period) and some 
indicator of magnetic activity, such as the unsigned average large-scale 
surface fields ($\langle |B_{\rm V}|\rangle$) or the fractional X-ray and 
H$\alpha$ luminosities, also known as coronal ($L_{\rm X}/L_{\rm bol}$) and 
chromospheric ($L_{{\rm H}\alpha}/L_{\rm bol}$) activity indicators, 
respectively. As shown by \citet{noyes84}, this relationship is better 
understood in terms of the Rossby number, $Ro$ (the ratio 
between rotation period, $P_{\rm rot}$, and the local convective turnover 
time, $\tau_{\rm c}$), than in terms of $P_{\rm rot}$. 
\citet{pizzolato03} showed for the first time that the 
rotation-magnetic activity relation was characterised by the existence 
of two distinct regions: the saturated region (formed by fast rotators 
- $P_{\rm rot}$$\leq$2~days) and 
the unsaturated region (composed by slow rotators - $P_{\rm rot}$$>$2~days). 
$L_{\rm X}/L_{\rm bol}$ (or another magnetic activity indicator) increases 
as $Ro$ decreases down to a saturation threshold value, $Ro_{\rm sat}$$\sim$0.1,
and remains constant at a saturation level, 
$(L_{\rm X}/L_{\rm bol})_{\rm sat}$$\sim$$10^{-3}$, for
$Ro$$<$$Ro_{\rm sat}$ \citep{wright11,wright18}. The causes of saturation are 
still unclear. 
This relation became 
a powerful tool to study stellar magnetism and is based on a general 
power-law function of the form $L_{\rm X}/L_{\rm bol} \propto Ro^{\beta}$,
where $\beta$ is a parameter to be adjusted with observations. Values of 
$\beta$ usually found in the literature are 
$\beta \!=\!-2$ \citep{pizzolato03}, 
$\beta \!=\!-2.7 \!\pm \!0.13$ \citep{wright11}, 
$\beta \!=\!-1.38 \! \pm \! 0.14$ \citep{vidotto14}, 
$\beta \!=\!-2.3^{+0.4}_{-0.6}$ \citep{wright18}, 
$\beta \!=\!-1.40 \! \pm \! 0.10$ \citep{see19} and 
$\beta \!=\!-2.4 \! \pm \! 0.1$ \citep{landin23}.

Describing the rotation-magnetic activity relation in terms of the Rossby 
number makes clear its connection with the stellar dynamo theory, as the 
dynamo number ($N_{\rm D}$, the efficiency of the dynamo in the mean-field dynamo theory) 
is proportional to the inverse 
square of the Rossby number ($N_{\rm D}\propto Ro^{-2}$). Consequently, 
the dynamo efficiency increases as $Ro$ decreases.
As $Ro$ plays an important role in the stellar magnetic activity studies,
determinations of $\tau_{\rm c}$ are of fundamental interest, since they 
cannot be directly measured. They can be obtained either semi-empirically
(\citealt{noyes84}, \citealt{pizzolato03}) or theoretically by 
stellar evolution models (\citealt{kim96}, \citealt{landin10}).
For a given stellar mass and age, the convective turnover time
varies significantly with the radial location. The location mostly used in 
the literature to determine local convective turnover times  
is one half a mixing length above the base of the convective zone 
\citep{noyes84}. This standard location coincides with the 
tachocline for partially convective stars, but it is not suitable for fully 
convective stars, in which there is no tachocline. In the Mixing Length Theory,
the adjustable mixing length parameter, $\ell$, is scaled with the pressure
scale height $H_{\rm p}$ as $\ell$=$\alpha H_{\rm p}$ ($\alpha$ is the
convection efficiency) and, 
during the fully convective phase, the modelled value of $H_{\rm p}$ at
the base of the convective zone is very high, and so is the mixing length.
Consequently, the standard location where $\tau_{\rm c}$ should be calculated
becomes larger than the stellar radius for fully convective configurations 
($M$$\leq$0.3\,M$_{\odot}$ at any evolutionary stage and 
$M$$>$$0.3\,{\rm M}_{\odot}$ before developing
the radiative core), making $\tau_{\rm c}$ calculations
unfeasible through this standard prescription. 
In order to overcome 
this problem, \citet{landin23} developed a method to obtain the location ($r$)
where to calculate $\tau_{\rm c}$ in terms of $H_{\rm P}$, 
allowing the theoretical determination of $\tau_{\rm c}$ of fully convective 
stars in a self consistent way with the traditional location prescribed by 
\citet{noyes84}.
For 10 selected ages in the range of 6.0$\leq$$\log(t/{\rm yr})$$\leq$10.14, 
\citet{landin23} analysed how the standard location $r/H_{\rm P}$ varied with stellar mass 
for partially convective stars, linearly fitted $r/H_{\rm P}$ as a function of the 
stellar mass and extrapolated it for fully convective stars. 
These linear fits of $r/H_{\rm p}$ as a function of mass were
introduced in the {\ttfamily ATON} stellar evolution code as alternative locations to the standard 
$\tau_{\rm c}$ calculation whenever $r$ is larger than the stellar radius. 
Otherwise, the standard location is adopted.

Most, if not all, low-mass pre-main sequence (pre-MS) stars exhibit some manifestation of magnetic
fields \citep{donati09}. As they miss a tachocline in the beginning of their pre-MS phase, the 
$\alpha$$-$${\mathrm \Omega}$ dynamo, which is an interface dynamo, is not supposed 
to be operating in this very early evolutionary phase. So, the observed magnetic activity in these stars should be 
produced by another kind of dynamo process, like a distributed dynamo, as
suggested by \citet{durney93}.
The characteristic shape of the rotation-activity relation 
exhibited by main sequence (MS) stars is not observed for pre-MS stars
\citep{flaccomio03}. The latter are seen only in the saturated region and show 
a considerable dispersion in magnetic activity levels \citep{preibisch05}. 
The fact that fully convective (pre-MS and MS) stars are preferably 
found in the saturated region, while solar-like (partially convective) stars 
are found in both regions, reinforces the idea that the dynamo operating 
in partially and fully convective stars are different. However, observations
by \citet{wright16} and \citet{wright18} indicate that partially and fully 
convective stars follow the same rotation-activity relation, implying that
they should operate very similar rotation-dependent dynamos in which the tachocline 
would not be a crucial ingredient.
\citet{landin23} present a theoretical and observational review 
of stellar magnetic activity in pre-MS and MS stars.

Low-mass stars in young clusters (ages$\leq$100\,Myr) are known to be very active and to exhibit strong magnetic
fields (specially those fast rotating), even though they do not follow 
the Skumanich law \citep[${\mathrm \Omega } \propto t^{-2}$,][]{alphenaar81}. 
For a few million years, their magnetic field lines are supposed to be anchored in a circumstellar disk,
formed during the 
star formation process, within a few stellar radii. The 
interaction between the stellar magnetic field and the disk regulates the 
star's angular velocity,
counteracting the tendency to spin up due to 
accretion of disk material of high specific angular momentum and 
to readjustments in the moment of inertia as the star contracts towards the MS. 
The magnetic torques acting on the central star during the disk 
lifetime transfer large amounts of its angular momentum to the disk and 
impose
the star to rotate with a virtually constant surface angular 
velocity. 

The observed rotation rates of stars in clusters of different ages 
suggest that the fast rotation phenomenon depends on mass. According to
several works, such as those of \citet{stauffer97}, \citet{prosser95} and 
\citet{stauffer87}, rapid rotation decreases slower for lower mass stars 
(the spin-down time scales expected for stars with $M$$<$0.5\,${\rm M_{\odot}}$
are longer than those for stars with $M$$>$1.0\,${\rm M_{\odot}}$).
\citet{attridge} and \citet{choi96} showed that T\,Tauri stars in the Orion 
Nebula Cluster have a very characteristic rotation period distribution: 
Classical T\,Tauri stars (CTTS), which still accrete from a disk, have a narrow period distribution
with a peak at about 8-10~days, while Weak-line T\,Tauri stars (WTTS), that no longer accrete from a disk, have 
a broader distribution showing only a tail of slow rotators.
In addition, WTTS appear to rotate faster, on average, than CTTS 
\citep{henderson2012}. 
Orion Nebula Cluster (ONC, 1\,Myr), NGC\,2264 (3.5\,Myr), IC\,348 (2.5\,Myr) 
and NGC\,2362 (3.3\,Myr) are some of the most studied young stellar clusters and
all of them exhibit bimodal and dichotomic rotation period distributions. The 
bimodality is related to the presence of two peaks in the period distribution and
the dichotomy is associated to the mass dependence of the distribution. 
Detailed studies of the rotation history of these clusters can be found
in the literature, for example: ONC was analysed by \citet{herbst02}, 
NGC\,2264 was investigated by \citet{lamm05}, 
\citet{cieza06} studied IC\,348, \citet{irwin08} examined NGC\,2362, the rotational properties
of ONC and NGC\,2264 was reviewed by \citet{landin16} and \citet{landin2021}
inspected such properties of IC\,348 and NGC\,2362. 
As the distribution of rotation periods in the pre-MS must evolve 
towards that of the MS, we have to be able to describe the process 
that leads from the wide distribution seen in the pre-MS to the even wider one 
seen in the MS. However, there is not a single model that manages to do this.
In order to reproduce the evolution of stars that reach the MS
with the highest rotation rates (near their breakup velocities), one should consider an 
evolution conserving angular momentum during all the pre-MS phase. On the other 
hand, to yield the rotation of stars spinning with a small fraction of 
their breakup velocities, one should consider that these stars were locked to their 
disks for a given time, keeping their angular velocities constant, and only 
after the locking phase they were released to spin up conserving angular 
momentum. \citet{landin16} present a theoretical and observational review 
of angular momentum evolution of pre-main sequence stars.

In this work, we use the {\ttfamily ATON} stellar evolution code to
investigate the influence of the disk-locking mechanism on local convective
turnover times and, consequently, on the Rossby numbers, and in the 
rotation-magnetic activity relation itself. Observational data of 
rotation periods, fractional X-ray 
luminosities and unsigned average large-scale surface magnetic fields from
\citet{vidotto14} are used to constrain our models.

In Section~\ref{models}, we briefly describe the {\ttfamily ATON} code and
the input parameters used in this work. 
Section~\ref{resDLtau} presents and discusses our results on convective 
turnover times and convective velocities obtained with disk-locking and 
angular momentum conserving models. Our new theoretical local convective
turnover times are compared to those existing in the literature and 
observational data are used to test our theoretical results in 
Section~\ref{applications}. Finally, Section~\ref{conclusions} presents our 
conclusions.

\section{Models and input physics} \label{models}

In the {\ttfamily ATON} code version used in this work, convection is treated
according to the traditional Mixing Length Theory \citep{bohmvitense58}, with the
parameter representing the convection efficiency set as $\alpha$$=$$2$. Surface boundary
conditions were obtained from non-grey atmosphere models \citep{allard00} with
a match between surface and interior at an optical depth of 10.  We used
the opacities reported by \citet{rogers1} and \citet{alexander} and the equations of
state from \citet{rogers2} and \citet{mihalas}. We assume that the elements
are mixed instantaneously in convective regions. Our tracks start from a fully
convective configuration with central temperatures in the range
$5.35$$<$$\log_{10}(T_{\rm c}/{\rm K})$$<$$5.70$,
follow deuterium and lithium burning, and end at the MS
configuration, as discussed in \citet{landin06}, who also presented a detailed discussion about the zero 
point ages of stellar models.

We generated two sets of models. In the first (hereafter AMC models), we considered conservation
of angular momentum throughout all stellar evolution. In the second set (hereafter DL models), we simulated 
the disk-locking mechanism, by considering an evolution with constant angular 
velocity during the first evolutionary stages, followed by conservation of angular momentum. 
In these models, the star-disk interaction treatment is not included in the code, and its 
effects on the stellar rotation are only mimicked by locking it for a while \citep{landin16}.

For AMC models, the initial angular momentum of each model was obtained 
according to the \citet{kawaler87} relation\footnote{For $M$$\ge$$0.2\,{\rm M}_{\odot}$, we
used the central values given by the \citet{kawaler87} relation to obtain the initial angular
momentum ($J_{\rm in}$) of our models. Due to non-convergence reasons, we used a slightly smaller value 
of $J_{\rm kaw}$ as $J_{\rm in}$ for 0.1\,M$_{\odot}$, namely 
$J_{\rm in}$$=$$7.943$$\times$$10^{48}~{\rm g~cm^2~s^{-1}}$, which corresponds to 
a rotation period of $\sim$30~days and is within the error bars in Eq.~\ref{eqkaw}.} 
\vspace{-0.7\baselineskip}
\begin{equation}
J_{\rm kaw}=(1.562 \pm 0.504)\times 10^{50} \left( {M \over M_{\odot}} \right) ^{(0.985 \pm 0.140)}
~~~{\mathrm{g~cm^2~s^{-1}.}}
\label{eqkaw}
\end{equation}
These initial angular momenta correspond to periods of $\sim$15~days for 
a 0.1\,${\rm M}_{\odot}$ model and $\sim$418~days for a 1.2\,${\rm M}_{\odot}$ 
model.

For DL simulating models, the initial angular momentum corresponds to the locking 
period, $P_{\rm lock}$, of $\sim$8\,days\footnote{Also due to non-convergence reasons, we used initial
angular momenta smaller than those corresponding to 8~days for models with $M$$\ge$$0.8\,{\rm M}_{\odot}$. 
The values used correspond to rotation periods of 8.5, 9.6, 10.6, 11.4 and 18.8~days
for 0.8, 0.9, 1.0, 1.1 and 1.2\,M$_{\odot}$, respectively.} based on the period distribution of 
CTTS in the ONC presented by \citet{herbst02}.
To investigate the effect of using different locking periods on $\tau_{\rm c}$,
we ran some additional DL models with $P_{\rm lock}$=400\,days. The disk 
lifetimes, $T_{\rm disk}$, used in this work are 1 and 3\,Myr, because
according to \citet{monsch23}, the majority of circumstellar disks dissipate after 
$\sim$1-3\,Myr.

The evolutionary tracks were computed in the mass range of 0.1 to 1.2\,M$_{\odot}$
(in 0.1\,M$_{\odot}$ steps). We adopt the solar chemical composition ($X$=$0.7155$ and
$Z$=$0.0142$) by \citet{asplund09}.  More details on the physics of the {\ttfamily
ATON} models are given by \citet{landin06,landin23}.

The current version of the {\ttfamily ATON} code allows for choosing among three
rotational schemes \citep{mendes99}, (1) rigid body rotation throughout the whole
star, (2) local conservation of angular momentum in the whole star (which leads to
differential rotation) and (3) local conservation of angular momentum in radiative
regions plus rigid body rotation in convective regions. 
Internal redistribution of angular momentum and angular momentum loss by
stellar magnetised winds are implemented in the {\ttfamily ATON} code only for 
rotational scheme 3. 
However, there is
observational evidence that the Sun's radiative core rotates as a solid body and the
convective envelope rotates differentially, opposite to scheme 3 \citep{thompson03}.
Here, our rotating models were generated according to scheme 1, because this is the
only rotational scheme for which disk-locking mechanism is implemented in the 
{\ttfamily ATON} code.
We leave for the future the implementation of a 4$^{th}$ rotational scheme with a
rotational profile closer to that of the Sun and disk-locking mechanism for
all rotational schemes.

\section{Theoretical results} \label{resDLtau}

By using the version of the {\ttfamily ATON} code described in \citet{landin23},
we followed the evolution of local convective turnover times from the pre-MS 
to the beginning of the MS for 0.1-1.2\,M$_{\odot}$ stars and tabulated them 
together with the corresponding evolutionary tracks. Table~\ref{tabtrack} presents
the 1\,M$_{\odot}$ DL model ($P_{\rm lock}\!\sim \!8$\,days and $T_{\rm disk}\!=\!3$\,Myr) 
as an example of such tables. 
For DL models, rotation periods start from the value given by the $P_{\rm lock}$ parameter,
are kept constant during the first million years (according to the $T_{\rm disk}$ parameter)
and then evolve considering conservation of angular momentum from this time on.
For models considering angular momentum conservation during all stages of evolution, the 
initial rotation periods are obtained from the initial angular momenta given in Eq.~\ref{eqkaw}.

\begin{table}[h]
\caption{Evolutionary track
for 1\,M$_{\odot}$ star (DL model with $P_{\rm lock}$$\sim$$8\,$days and 
$T_{\rm disk}$=$3\,$Myr\tablefootmark{a}). 
}
\label{tabtrack}
\centering
{\small
\advance\tabcolsep by -3.0pt
\begin{tabular}{crrclrrcl}
\hline \hline \\ [-11pt]
${{\log\!\frac{t}{\rm yr}}}\vphantom{\Big|}$\,\, & $\log\!\frac{L}{L_{\odot}}$ & ${\log}\atop{(T_{\rm eff}/{\rm K)}}$ & ${\log}\atop{(g/{\mathrm{cm\,s^{-2}}})}$  & ${{\log}\atop{(\tau_{\rm c}/{\rm d)}}}\vphantom{\Big|}$ & ${{\log}\atop{(\tau_{\rm g}/{\rm d})}}$ & ${P_{\rm rot}/{\rm d}}$ & $Ro$ &  ${M_{\rm rad}}\atop{(M_{\odot})}$ \\ [-0pt] \hline  
	&   &  &  &  &  &  &  & \\ [-08pt]
         
  2.233 &  1.665 & 3.600 & 2.127 & 1.976 & \,2.910 & 10.580 & 0.112 & 0.000 \\ [-1.7pt]
  3.668 &  1.601 & 3.623 & 2.280 & 1.901 & \,2.805 & 10.580 & 0.133 & 0.000 \\ [-1.7pt]
  4.203 &  1.382 & 3.634 & 2.545 & 1.892 & \,2.792 & 10.580 & 0.136 & 0.000 \\ [-1.7pt]
  4.643 &  1.163 & 3.642 & 2.795 & 1.874 & \,2.708 & 10.580 & 0.141 & 0.000 \\ [-1.7pt]
  5.186 &  0.989 & 3.647 & 2.989 & 1.871 & \,2.687 & 10.580 & 0.143 & 0.000 \\ [-1.7pt]
  5.352 &  0.836 & 3.651 & 3.158 & 1.870 & \,2.781 & 10.580 & 0.143 & 0.000 \\ [-1.7pt]
  5.606 &  0.615 & 3.653 & 3.389 & 1.869 & \,2.785 & 10.580 & 0.143 & 0.000 \\ [-1.7pt]
  5.882 &  0.394 & 3.652 & 3.604 & 1.873 & \,2.810 & 10.580 & 0.142 & 0.000 \\ [-1.7pt]
  6.164 &  0.173 & 3.646 & 3.802 & 1.942 & \,2.853 & 10.580 & 0.121 & 0.003 \\ [-1.7pt]
  6.454 & -0.048 & 3.637 & 3.987 & 2.144 & \,2.589 & 10.580 & 0.076 & 0.160 \\ [-1.7pt]
  6.779 & -0.253 & 3.632 & 4.173 & 2.059 & \,2.406 &  6.859 & 0.060 & 0.428 \\ [-1.7pt]
  7.054 & -0.316 & 3.650 & 4.305 & 1.893 & \,2.211 &  4.469 & 0.057 & 0.680 \\ [-1.7pt]
  7.248 & -0.221 & 3.689 & 4.367 & 1.683 & \,1.990 &  3.020 & 0.063 & 0.853 \\ [-1.7pt]
  7.372 & -0.049 & 3.734 & 4.376 & 1.439 & \,1.756 &  2.082 & 0.076 & 0.946 \\ [-1.7pt]
  7.478 &  0.019 & 3.766 & 4.437 & 1.177 & \,1.504 &  1.418 & 0.094 & 0.983 \\ [-1.7pt]
  8.563 & -0.092 & 3.759 & 4.520 & 1.213 & \,1.521 &  1.354 & 0.083 & 0.980 \\ [-1.7pt]
  9.277 & -0.043 & 3.764 & 4.487 & 1.187 & \,1.500 &  1.363 & 0.089 & 0.982 \\ [-1.7pt]
  9.490 & -0.001 & 3.767 & 4.458 & 1.178 & \,1.481 &  1.372 & 0.091 & 0.984 \\ [-1.7pt]
  9.624 &  0.040 & 3.769 & 4.427 & 1.150 & \,1.460 &  1.389 & 0.098 & 0.985 \\ [-1.7pt]
  9.743 &  0.094 & 3.772 & 4.384 & 1.142 & \,1.443 &  1.423 & 0.103 & 0.986 \\ [-1pt] \hline
\end{tabular}
\tablefoot{Column\,1 gives the $\mathrm{log}_{10}$ of stellar 
age; col.\,2 the $\mathrm{log}_{10}$ of bolometric luminosity; col.\,3 the $\mathrm{log}_{10}$ 
of effective temperature; col.\,4 the $\mathrm{log}_{10}$ of effective gravity; col.\,5 the 
$\mathrm{log}_{10}$ of local convective turnover time; col.\,6 the $\mathrm{log}_{10}$ of 
global convective turnover time; col.\,7 the rotation period; col.\,8 the Rossby number; and 
col.\,9 the radiative core mass. Throughout this work all the logarithms are in base 10.\\ 
\tablefoottext{a}{The complete version of the table, including 24 tracks for masses in the range 
0.1-1.2\,M$_{\odot}$ (in 0.1\,M$_{\odot}$ increments) for DL and AMC models, is available at 
the CDS.}}
}
\end{table}

The initial angular velocities of most of our DL models
are higher than those of AMC models.
However, the angular velocity,
${\mathrm \Omega}$, is kept constant during 3\,Myr in the DL models, while in the
AMC models the stars are free to spin up since the
beginning of their evolutions. As a consequence, AMC models 
become faster than DL models in the first million years
of evolution, reaching the ZAMS with higher angular velocities (for 
1.0\,M$_{\odot}$, for example, ${\mathrm \Omega}_{\rm AMC}$$\approx$4${\mathrm \Omega}_{\rm DL}$ at the ZAMS).

As can be seen in Table~\ref{tabtrack}, our 1\,M$_{\odot}$ model 
at the solar age shows a rotation period of 1.389\,days, which is quite different from the current Sun. This is mainly due to
the fact that we did not take into account some physical phenomena such as angular momentum loss
by stellar winds. Such phenomena have large impact on the rotation rate, which in turn has
a secondary effect on local convective turnover times, that are mainly determined by the 
mass of the model.
According to our 1\,M$_{\odot}$ DL model, at the age of the Sun, the solar local convective turnover time is 
$\tau_{\rm c,\odot}$$=$14.13\,days, which corresponds to $Ro_{\odot}$$=$1.77 by using 
$P_{\rm rot,\odot}$$=$25\,days \citep[the same value used by][]{vidotto14}. 
This value is consistent to that found semi-empirically by \citet{pizzolato03}, which 
is $\tau_{\rm c,\odot}$$=$12.59\,days ($Ro_{\odot}$$=$1.99). 

\subsection{Convective velocities calculations}\label{velc}

\begin{figure}
\centering{
\includegraphics[width=09.2cm]{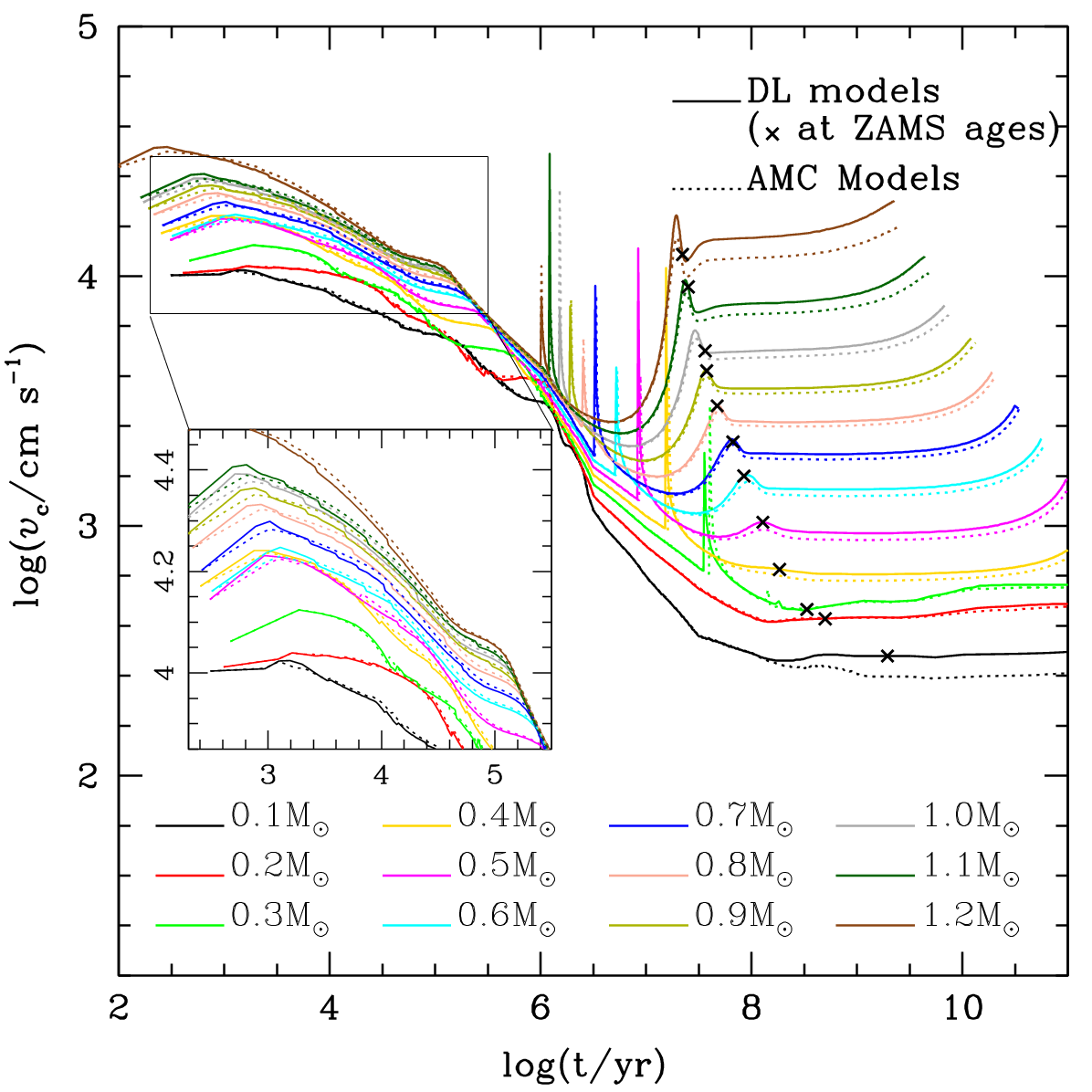}
\caption{Convective velocity as a function of age 
for different sets of models (DL models are drawn in solid lines 
and AMC models are shown in dotted lines). The curves referring to each 
stellar mass are drawn in a different colour according to the legend. 
Crosses ($\times$) show the ZAMS ages for each mass model.
The inset shows in detail the temporal evolution of 
$\varv_{\rm c}$ during the beginning of the pre-MS phase.}
\label{velcage}
}
\end{figure}

In the framework of the Mixing Length Theory, which provides a description of the
average behaviour of convective motions, the convective velocity, $\varv_{\rm c}$, 
of the rising or falling material is related to its excess or deficit of temperature.
For a given stellar mass and age, $\varv_{\rm c}$ increases radially outwards in 
the star \citep[see Fig.\,6 of][]{landin23}, and its value of greatest interest for 
this work is that used in the calculation of $\tau_{\rm c}$.
Fig.~\ref{velcage} shows plots of convective velocities, which will be used 
for this purpose, as a function of age, for masses in the range of
0.1$-$1.2\,${\rm M_{\odot}}$ with 0.1\,M$_{\odot}$ increments. The convective
velocities shown in Fig.~\ref{velcage} were evaluated for $\tau_{\rm c}$ calculation 
at the traditional location, at one-half of a mixing length above 
the base of the convective zone, for partially convective configurations, while for fully convective
configurations, $\varv_{\rm c}$ was estimated at the alternative location determined 
by \citet{landin23}.

Values of $\varv_{\rm c}$ decrease during early phases of the
pre-MS, increase after the formation of the radiative core and remain nearly constant during the MS;
the higher the stellar mass, the higher the convective velocity. In the
initial part of the pre-MS (3$\lesssim$$\log(t/yr)$$\lesssim$6),
$\varv_{\rm c}$ is slightly lower for models simulating DL than for AMC models,
while during the MS, models simulating DL yield
higher convective velocities.
This change in the $\varv_{\rm c}$ behaviour happens around
1\,Myr, when AMC models become faster rotators than DL models. 

\begin{figure}
\centering{
\includegraphics[width=0.45\textwidth]{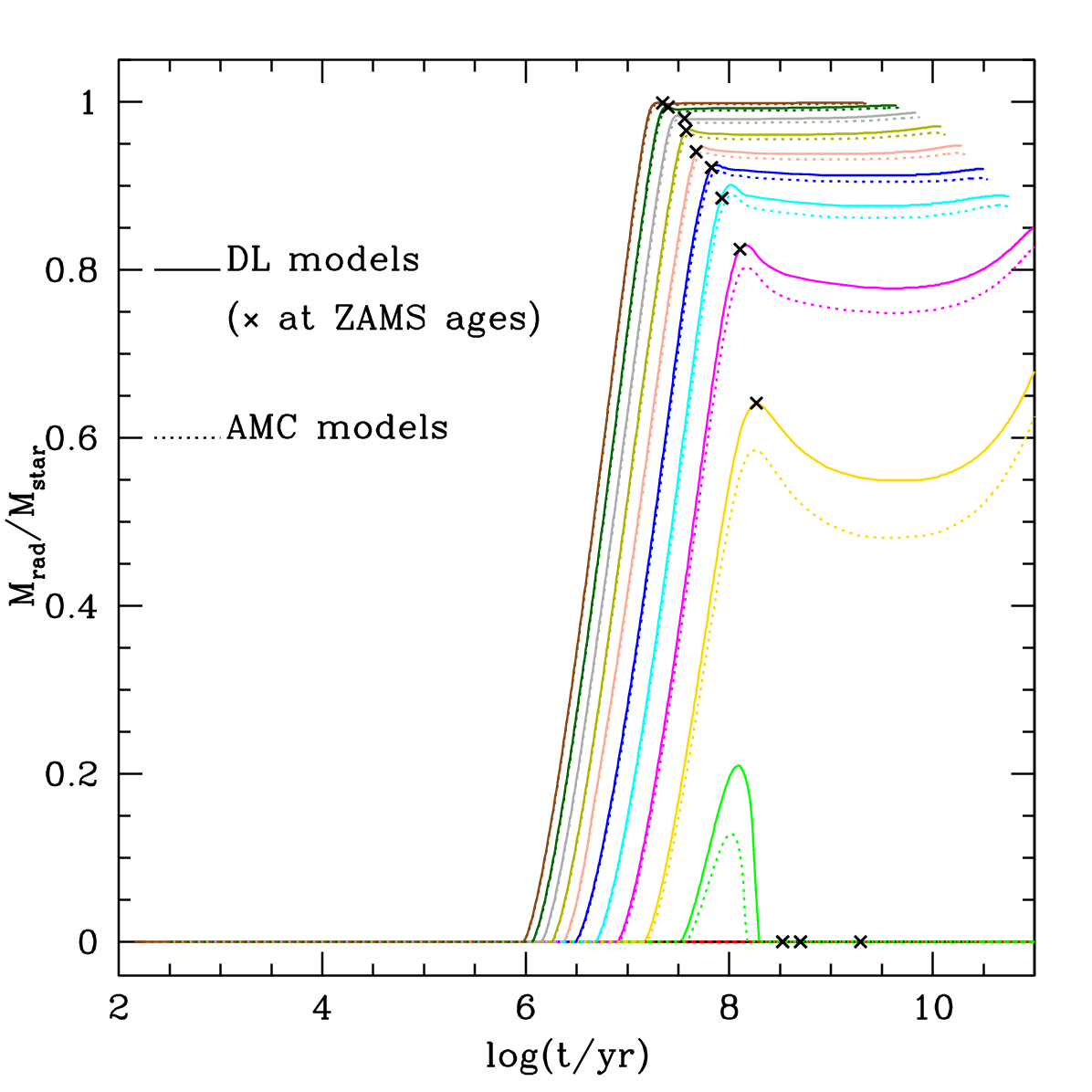}
\caption{Fractional mass of the radiative core for DL and
AMC models. Symbols and colours are the same as in Fig.~\ref{velcage}.} 
\label{mrradxage}
}
\end{figure}

The $\varv_{\rm c}$ behaviour can be understood by means of an extension of 
the mass-lowering effect \citep{sackmann70}, which states that rotating stars 
mimic non-rotating stars with smaller masses.
Here, we are extending the mass-lowering effect to stars with different rotation 
rates, in which stars with higher rotation rates mimic non-rotating stars with 
even lower masses than stars with lower rotation rates do. 
The fractional mass of the radiative core of each 
stellar mass for AMC and DL models, shown in Fig.~\ref{mrradxage}, seem to corroborate our ideas. 
AMC models produce smaller fractional radiative core masses than
DL models, as if they were models with smaller masses.

A consequence from the behaviour seen in
Fig.~\ref{mrradxage} is that including disk-locking in
the models results in weaker effects of rotation in the MS. During the
disk-locking phase, stars contract keeping their angular
velocities constant until the dissipation of the disk ($\sim$3\,Myr), when
they start spinning-up conserving angular momentum, resulting in slower rotation
rates in the MS, as compared with AMC models, that evolve without any locking.
So, the faster the rotation, the stronger the mass-lowering effect and
the slower the convective velocities.

In the MS the differences in $\varv_{\rm c}$ are higher for more massive stars, 
except for the 0.1\,M$_{\odot}$ model whose difference in convective velocity 
obtained by the two sets of models is comparable to that obtained by the 
1.2\,M$_{\odot}$ model.

\begin{figure}
\centering{
\includegraphics[width=08.5cm]{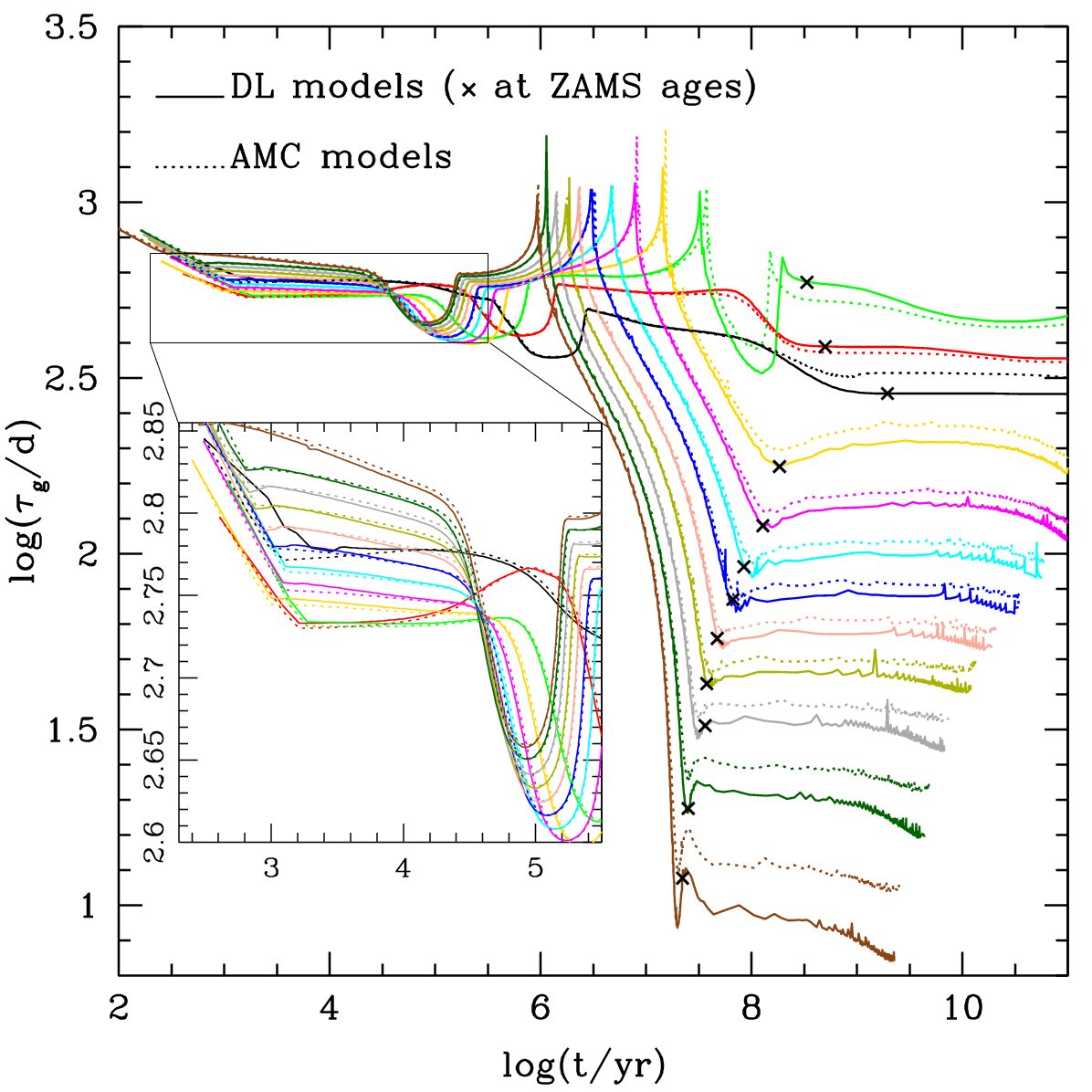}
\caption{Global convective turnover time as a function of age and stellar mass
for different sets of models. The inset shows in detail the time
evolution of $\tau_{\rm g}$ during the beginning of the pre-MS 
phase. Colours and symbols are the same as in Fig.~\ref{velcage}.} 
\label{taugage}
}
\end{figure}

\subsection{Global convective turnover times calculations}\label{taug}

The global (or non-local) convective turnover time, $\tau_{\rm g}$, is defined as

\vspace{-1.5\baselineskip}

\begin{equation}
\tau_{\rm g}=\!\int_{R_{\rm b}}^{R_{\rm star}}\varv_{\rm c}^{-1}{\rm d}r,
\label{taugdefeq}
\end{equation}

\vspace{-0.35\baselineskip}

\noindent
where $R_{\rm star}$ is the stellar radius and $R_{\rm b}$ is the radial position of the base of
the convective zone, the same as that of the tachocline for
partially convective stars; $R_{\rm b}$=0 (stellar centre) for fully
convective stars.
According to \citet{kim96}, $\tau_{\rm g}$ is the characteristic timescale of
the convective overturn. It can be used to depict convection features of the whole stellar
convective region at each evolutionary stage, representing
an average convective timescale over the entire convective zone. In the absence of a
suitable prescription to calculate local convective turnover times of fully convective
stars, $\tau_{\rm g}$ is used to evaluate their Rossby numbers
instead of $\tau_{\rm c}$ \citep{irving23}.
Fig.~\ref{taugage} shows the global convective turnover times produced by AMC and
DL models as a function of age and mass. 
The behaviour of $\tau_{\rm g}$ is a consequence of its definition (Eq.~\ref{taugdefeq}),
which depends directly on the actual extension of the convective zone ($r_{\rm conv}$), 
but inversely on the convective velocity. 
For a star of a given mass and age, various combinations of these two effects 
can yield higher or lower $\tau_{\rm g}$, depending on which effect is dominant.

\begin{figure}
\centering{
\includegraphics[width=0.45\textwidth]{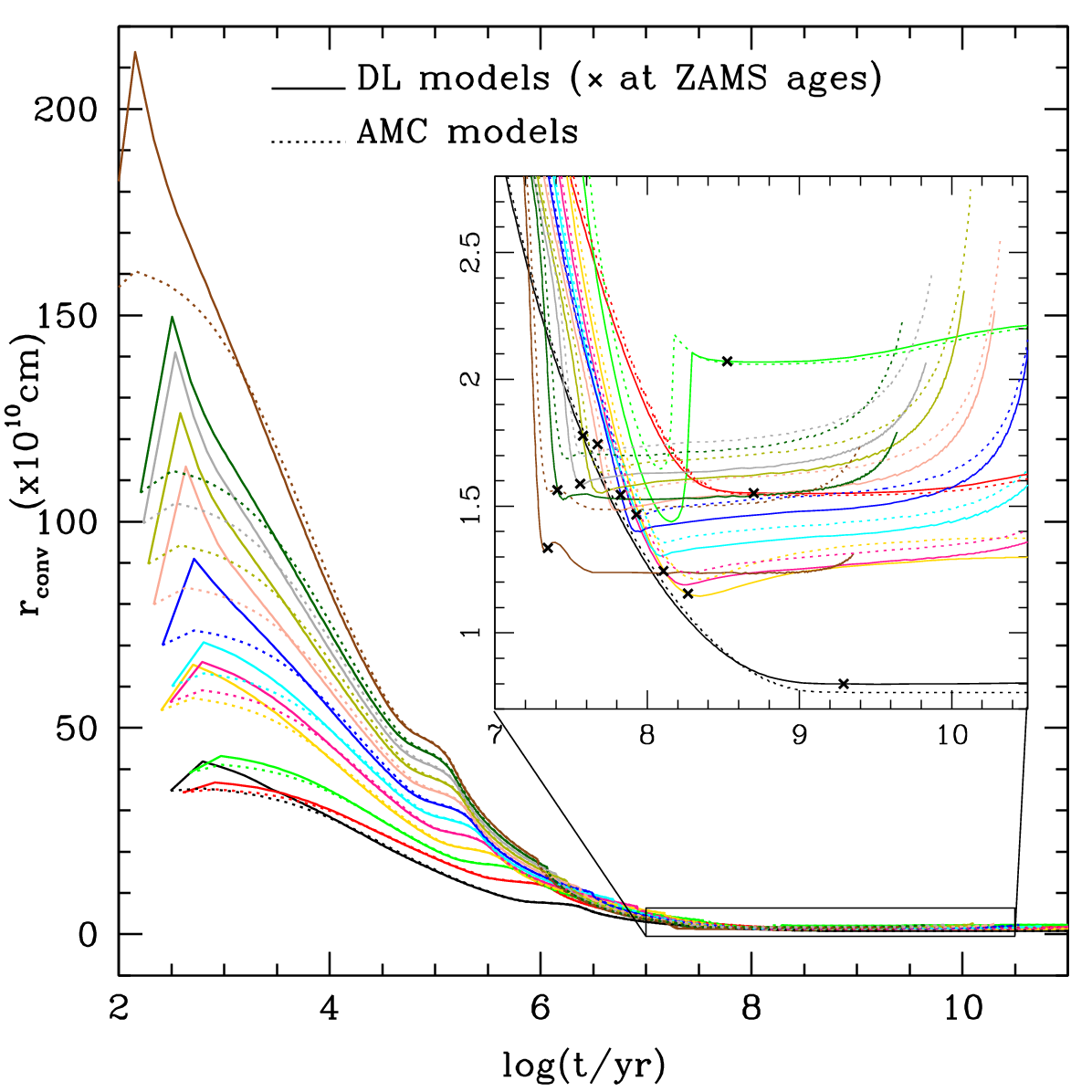}
\caption{Extension of the convective envelope in absolute terms (from the 
base to the top) as a function of mass and age for DL and AMC models. 
The inset shows in detail the time evolution of $r_{\rm conv}$ during the 
MS phase. 
Symbols and colours are the same as in Fig.~\ref{velcage}.} 
\label{rconv}
}
\end{figure}

The behaviour of $v_{\rm c}$ was already discussed in Section~\ref{velc} and 
before getting into the analysis of $\tau_{\rm g}$, we will describe the behaviour 
of the extension of the convective zone in absolute terms. Fig.~\ref{rconv} shows how $r_{\rm conv}$
varies with mass, age and model used. In the pre-MS (to the left of the crosses defining
the ZAMS), $r_{\rm conv}$ increases with mass and decreases with the stellar age. In the
very early phase of evolution (up to $10^3$ to $10^4$ years, depending on the 
stellar mass), $r_{\rm conv}$ produced by AMC models is smaller than that yielded
by DL models. After that age (and up to the ZAMS), the behaviour of $r_{\rm conv}$ 
is reversed.
In the MS (to the right of the crosses in Fig.~\ref{rconv}), $r_{\rm conv}$ shows small
changes with age for a given mass, but varies in a more complex way with mass. It sharply increases with 
mass from 0.1 to 0.3\,M$_{\odot}$, decreases abruptly 
from 0.3 to 0.4\,M$_{\odot}$ (because the star develops a radiative core), then gradually 
increases from 0.4 to 1.0\,M$_{\odot}$, reaches a local maximum (smaller than that 
for 0.3\,M$_{\odot}$), and decreases again from 1.0 to 1.2\,M$_{\odot}$. 
According to the model used, DL models produce $r_{\rm conv}$ larger than AMC
models for MS stars with $M\!\leq\!0.3\,\mathrm{M}_{\odot}$, while for MS stars with
$M\!>\!0.3\,\mathrm{M}_{\odot}$ AMC models yield larger $r_{\rm conv}$
than DL models do.

As well as \citet{landin23}, we noticed that the time evolution of $\tau_{\rm g}$,
showed in Fig.~\ref{taugage},
is different for fully and partially convective stars. 
For very low-mass stars ($M$$\le$0.3\,M$_{\odot}$), one can see only small 
variations (a factor of 3.5) on global convective turnover times, while variations 
reach almost 4 orders of magnitude for stars with $M\!>\!0.3\,\mathrm{M}_{\odot}$.

For $M\!\!\leq\!0.3\,\mathrm{M}_{\odot}$, the global convective turnover time 
does not vary significantly with age during the pre-MS and MS phases of evolution 
(less than an order of magnitude). In the beginning of
the pre-MS ($t\! \lesssim\!10^6$ years), $\tau_{\rm g}$
decreases with increasing mass, but in the MS the opposite way of behaving is observed. 
This is due to the behaviour of $\varv_{\rm c}$ and $r_{\rm conv}$
(Figs.~\ref{velcage} and \ref{rconv}, respectively), 
which essentially increase with mass for all ages, so that the former contributes to decrease 
$\tau_{\rm g}$ and the latter contributes to increase it. In the pre-MS, the dominant
effect on $\tau_{\rm g}$ is that of $\varv_{\rm c}$ and in the MS 
the dominant effect is that provided by $r_{\rm conv}$.

For $M\!\!>\!0.3\,\mathrm{M}_{\odot}$, $\tau_{\rm g}$ values do not change 
significantly with age for a given mass in the beginning of the pre-MS phase, in which stars are fully convective, 
then decrease when the radiative core is formed and remain roughly constant during the 
MS phase.
In the pre-MS, the higher the stellar mass the higher $\tau_{\rm g}$. As can be seen
in Figs.~\ref{velcage} and \ref{rconv}, both $\varv_{\rm c}$ and $r_{\rm conv}$ basically
increase with mass for all ages, such that the effect of $\varv_{\rm c}$ leads to a decrease 
in $\tau_{\rm g}$, while $r_{\rm conv}$ causes an increase in $\tau_{\rm g}$. In this
initial evolutionary phase, the influence of $r_{\rm conv}$ dominates that of $\varv_{\rm c}$
in the behaviour of $\tau_{\rm g}$. In the MS, $\tau_{\rm g}$ increases with
decreasing mass. Across this entire mass range, $\varv_{\rm c}$ increases with the stellar mass,
contributing to decrease $\tau_{\rm g}$, while $r_{\rm conv}$ has a more complex behaviour.
In the mass ranges 0.3-0.4\,M$_{\odot}$ and 1.0-1.2\,M$_{\odot}$, $r_{\rm conv}$ decreases 
with mass, reinforcing the effect of $\varv_{\rm c}$ and contributing to reduce $\tau_{\rm g}$ 
with mass, and in the range 0.4-1.0\,M$_{\odot}$, $r_{\rm conv}$ increases with mass, opposing 
the effect of $\varv_{\rm c}$ and contributing to increase $\tau_{\rm g}$ with mass.
The combination of these two effects results in the prevalence of $\varv_{\rm c}$ in the
behaviour of $\tau_{\rm g}$, making $\tau_{\rm g}$ decreases with increasing 
mass.

In our calculated models, the behaviour of $\tau_{\rm g}$ depends on the stellar mass.
In the beginning of the pre-MS, the global convective turnover time is practically
model independent.
For M$\geq$0.8\,M$_{\odot}$, AMC models generate $\tau_{\rm g}$ slightly longer than 
DL models, while for M$<$0.7\,M$_{\odot}$, the opposite trend is observed. 
It seems that 0.7\,M$_{\odot}$ is
a transition mass for this behaviour. In the MS, $\tau_{\rm g}$ is shorter for DL
models in comparison with AMC ones ($\tau_{\rm g,DL}$$<$$\tau_{\rm g,AMC}$),
except for 0.2 and 0.3\,M$_{\odot}$.
For $M\!\leq\!0.3\,\mathrm{M}_{\odot}$, DL models produce larger convective zones than AMC models
($r_{\rm conv,DL}\!\!>\!\!r_{\rm conv,AMC}$), favouring $\tau_{\rm g,DL}\!>\!\tau_{\rm g,AMC}$, 
while the convective velocities yielded by DL models are larger than those produced
by AMC models ($\varv_{\rm c,DL}\!>\!\varv_{\rm c,AMC}$), favouring 
$\tau_{\rm g,DL}<\tau_{\rm g,AMC}$.
The interplay between these two effects is dominated by the size of the convective zone for 
0.2 and 0.3\,M$_{\odot}$ (producing $\tau_{\rm g,DL}\!>\!\tau_{\rm g,AMC}$) and 
by the convective velocity for 0.1\,M$_{\odot}$ (yielding $\tau_{\rm g,DL}\!<\!\tau_{\rm g,AMC}$).
Among the fully convective models
($M\!\leq\!0.3\,\mathrm{M}_{\odot}$), one sees that the behaviour of DL and AMC
models for 0.1\,M$_{\odot}$ is the opposite of those of 0.2 and
0.3\,M$_{\odot}$. This can be explained by the fact that $\varv_{\rm c}$ (Fig.~\ref{velcage})
produced by AMC models for 0.2 and 0.3\,M$_{\odot}$ are only slightly lower
than those yielded by DL models while, for 0.1\,M$_{\odot}$, AMC models produce 
$\varv_{\rm c}$ noticeably lower than DL models do. 
Analyses with finer 
mass grid models (0.085, 0.09, 0.11, 0.12, 0.13, 0.14 and 0.15\,M$_{\odot}$, not shown 
here) reveal that such a switch in $\tau_{\rm g}$ behaviour occurs between 0.10 and 
0.11\,M$_{\odot}$ and $\tau_{\rm g,DL}$ continues
smaller than $\tau_{\rm g,AMC}$ for masses smaller than 0.10\,M$_{\odot}$
(see the following quantitative comparisons between DL and AMC values of $\tau_{\rm g}$ at 10\,Gyr:
$\tau_{\rm g,DL}/\tau_{\rm g,AMC}$ is 1.0251 for 0.2\,M$_{\odot}$, 1.0041 
for 0.11\,M$_{\odot}$, 0.8765 for 0.10\,M$_{\odot}$, 0.8423 for 0.09\,M$_{\odot}$ and
0.9102 for 0.085\,M$_{\odot}$).
As the additional AMC models also present convergence issues when adopting the central 
values of $J_{\rm Kaw}$ as $J_{\rm in}$, we used smaller input values which vary from 
$0.987\,J_{\rm Kaw}$ for 0.085\,M$_{\odot}$ to $0.997\,J_{\rm Kaw}$ for 0.15\,M$_{\odot}$.
According to our AMC models, $\tau_{\rm g}$ decreases with increasing $J_{\rm in}$. If it
were not for some convergence difficulties and had we used $J_{\rm in}=J_{\rm Kaw}$ for models 
with $M\!\leq\!0.15\,\mathrm{M}_{\odot}$, we would have obtained $\tau_{\rm g,DL}$ even smaller 
than $\tau_{\rm g,AMC}$, making the change in $\tau_{\rm g}$ behaviour near 0.1\,M$_{\odot}$ 
even more pronounced.
For $M\!>\!0.3\,\mathrm{M}_{\odot}$, the behaviour of $\tau_{\rm g}$ 
depending on the model used can be explained in a much simpler way, given that 
$r_{\rm conv,DL}\!<\!r_{\rm conv,AMC}$ 
and $\varv_{\rm c,DL}\!>\!\varv_{\rm c,AMC}$, both effects contribute 
to generate $\tau_{\rm g,DL}\!<\!\tau_{\rm g,AMC}$.

\begin{figure}
\centering{
\includegraphics[width=08.5cm]{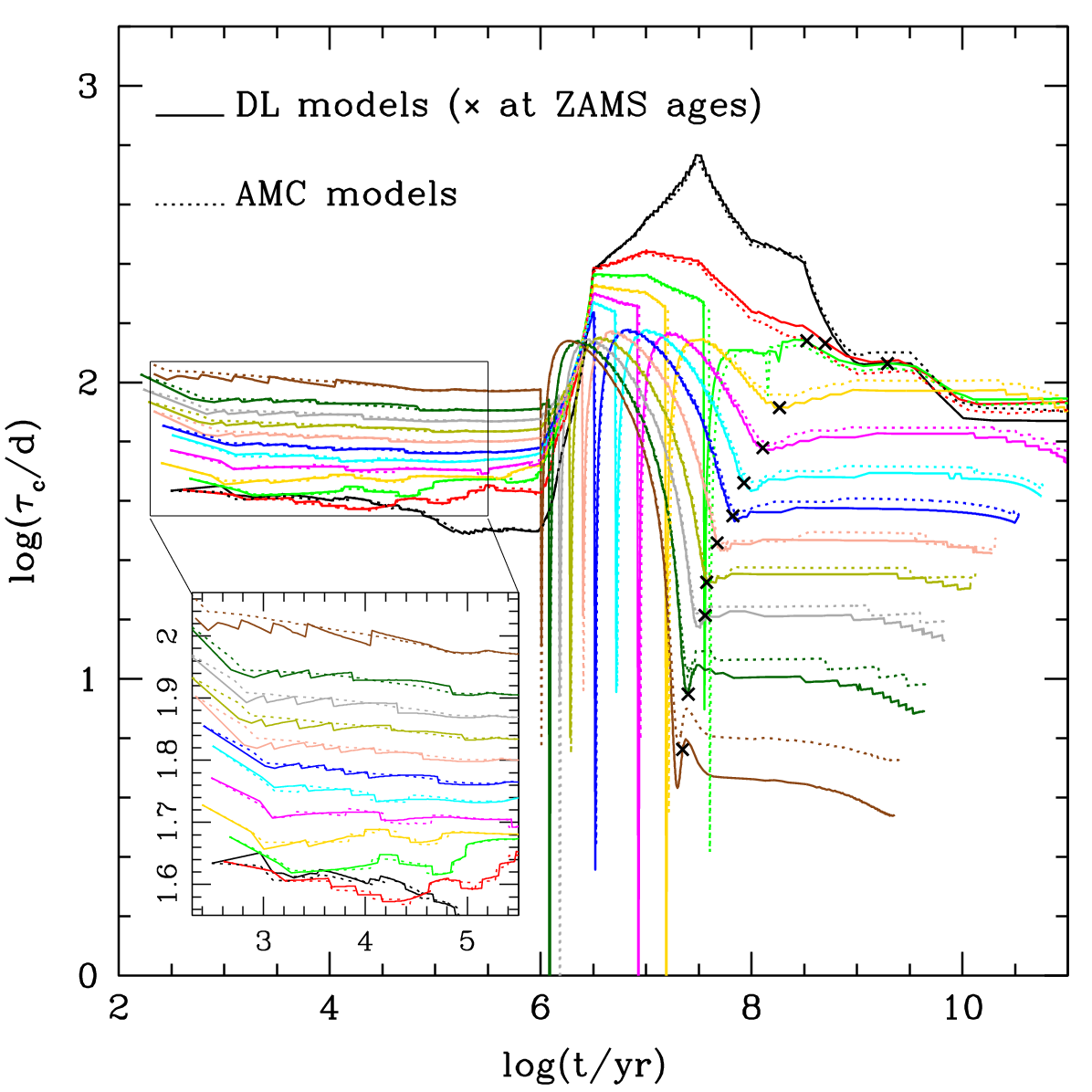}
\caption{Local convective turnover time as a function of age and stellar mass
for different sets of models. Symbols and colours have the same meanings as 
in Fig.~\ref{velcage}. The inset shows in detail the time evolution of 
$\tau_{\rm c}$ during the beginning of the pre-MS phase.}
\label{taulage}
}
\end{figure}

\subsection{Local convective turnover times calculations}\label{tauc}

The characteristic convective overturn timescale used to compute Rossby numbers
is the local convective turnover time, which differs from $\tau_{\rm g}$
only by a factor.
It is defined as $\tau_{\rm c}$$=$${\ell}/\varv_{\rm c}$ and is evaluated in deep
regions of the convective envelope,
where dynamo generation of magnetic fields is supposed to take place \citep{kim96}. 
Fig.~\ref{taulage} shows
local convective turnover times as a function of age and mass.
For partially convective configurations,
$\tau_{\rm c}$ was calculated at the standard location, $r$, one-half of a mixing length
above the base of the convective zone. For fully convective configurations (whose $r$ 
are larger than the stellar radii), $\tau_{\rm c}$ was calculated at an alternative
place related to $H_{\rm p}$, as described in \citet{landin23}.
$\tau_{\rm c}$ behaves roughly as $\tau_{\rm g}$, both as
a function of age and as a function of mass, with some differences in the low mass regime and in 
the range of 1-11\,Myr, when $\tau_{\rm c}$ produced by models with $M$$\leq$$0.3\,{\rm M}_{\odot}$ 
become smaller than those for 0.4\,${\rm M}_{\odot}$
models, as opposed to what happens regarding $\tau_{\rm g}$.
This is 
probably due to the method used to calculate $\tau_{\rm c}$ for fully convective 
stars, based on different linear fits for different age intervals and
excluding higher masses because they deviate more from the linear behaviour; see
\citet{landin23} for more details.
In the pre-MS, 
$\tau_{\rm c}$
is approximately constant 
and
does not show a strong dependence with modelling 
(differences are around 1\%). AMC models 
tend
to produce
higher $\tau_{\rm c}$ for higher masses ($M$$>$$0.7\,{\rm M}_{\odot}$) and DL
models tend
to yield
higher $\tau_{\rm c}$ for smaller masses ($M$$<$$0.6\,{\rm M}_{\odot}$).
As the stellar age increases, the cumulative effects of evolving with a higher 
angular velocity reflect in the convective properties of the stars and the 
differences in $\tau_{\rm c}$ produced by the two sets of models reach 8-10\% for 
ages greater than 100\,Myr. Stars with $M$$\ge$$1\,{\rm M_{\odot}}$ present the 
highest differences.
In the MS, values of $\tau_{\rm c}$ generated with models simulating the DL mechanism 
are shorter than those obtained with AMC models, except for 0.2 and 0.3M\,$_{\odot}$, repeating the previously described behaviour of $\tau_{\rm g}$. 

\begin{figure}
\centering{
\includegraphics[width=0.49\textwidth]{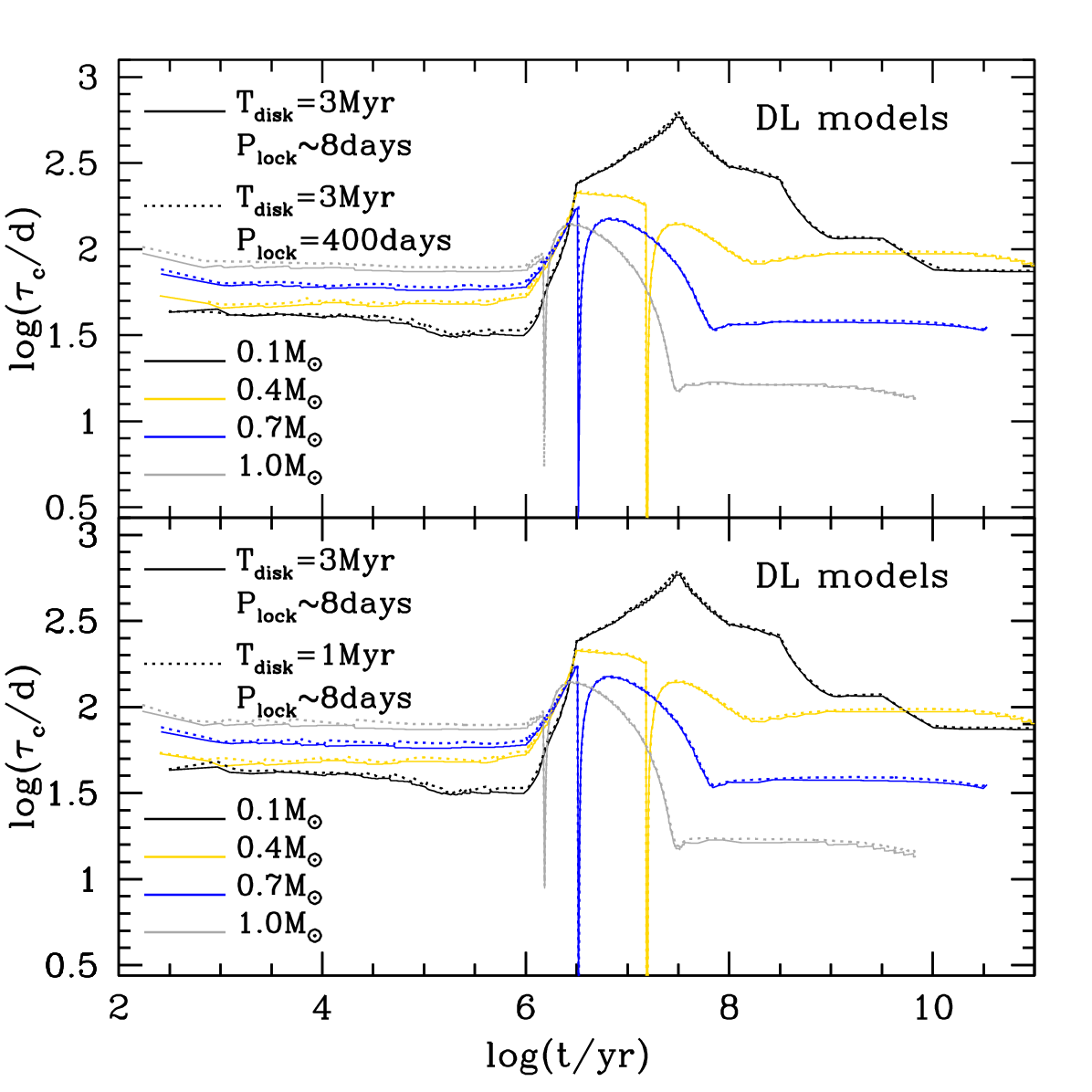}
\caption{Local convective turnover time as a function of age and mass for DL models with
different locking periods (top panel) and different disk lifetimes 
(bottom panel). Colours are the same as in Fig.\,\ref{velcage}.} 
\label{taulagediftdpl}
}
\end{figure}

The top panel of Figure \ref{taulagediftdpl} shows the local convective
turnover time as a function of age and mass produced by DL models 
with different locking periods ($P_{\rm lock}$$\sim$8\,days and 
$P_{\rm lock}$=400\,days, keeping the same $T_{\rm
disk}$=3\,Myr). For clarity, we show plots only for 4 stellar mass 
models, 0.1, 0.4, 0.7 and 1.0\,${\rm M}_{\odot}$. DL models with 
longer locking periods produce slightly longer values of local convective 
turnover times (except for the 1.0\,${\rm M}_{\odot}$ at the MS, for which
$\tau_{\rm c}$ is almost model independent). The reason is that DL models with
$P_{\rm lock}$=400\,days have larger mixing lengths and convective 
velocities than those with 
$P_{\rm lock}$=8\,days, that contribute to increase and decrease $\tau_{\rm c}$, respectively, 
and the dependence on $\ell$ dominates over that on $\varv_{\rm c}$.
The bottom panel of Fig.~\ref{taulagediftdpl} shows the local convective
turnover time as a function of age and mass produced by DL models with 
different disk lifetimes ($T_{\rm disk}$$=$1 and 3\,Myr,
keeping the same $P_{\rm lock}$$\sim$8\,days). We again show plots 
only for 4 stellar mass models, 0.1, 0.4, 0.7 and 1.0\,${\rm M}_{\odot}$. For 
all evolutionary phases and masses, DL models with $T_{\rm disk}$=1\,Myr 
yielded slightly longer $\tau_{\rm c}$ than DL models with $T_{\rm disk}$=3\,Myr, 
because, while $\ell$ is almost independent on T$_{\rm disk}$, DL models
with $T_{\rm disk}$=1\,Myr produce convective velocities slightly lower than those with $T_{\rm disk}$=3\,Myr,
since $\tau_{\rm c} \propto 1/ \varv_{\rm c}$.

$\tau_{\rm g}$ and $\tau_{\rm c}$ are quantities that cannot be directly observed
and the best way to estimate them is through stellar evolutionary models. They are extensively
used to determine Rossby numbers of stars operating different types of dynamos. $\tau_{\rm c}$
is employed in stars that have tachocline-based dynamos and, when there is no suitable
prescriptions to obtain it for stars that have no tachocline and harbour distributed dynamos, 
$\tau_{\rm g}$ is used.
$\tau_{\rm g}$ and $\tau_{\rm c}$ are mainly determined by the stellar mass, but 
non-standard physical ingredients, like the rotation rate, the way angular momentum evolves and
disk-locking parameters, have secondary effects, as discussed in Sections~\ref{taug} and \ref{tauc}.

\begin{figure}
\centering{
\includegraphics[width=0.495\textwidth]{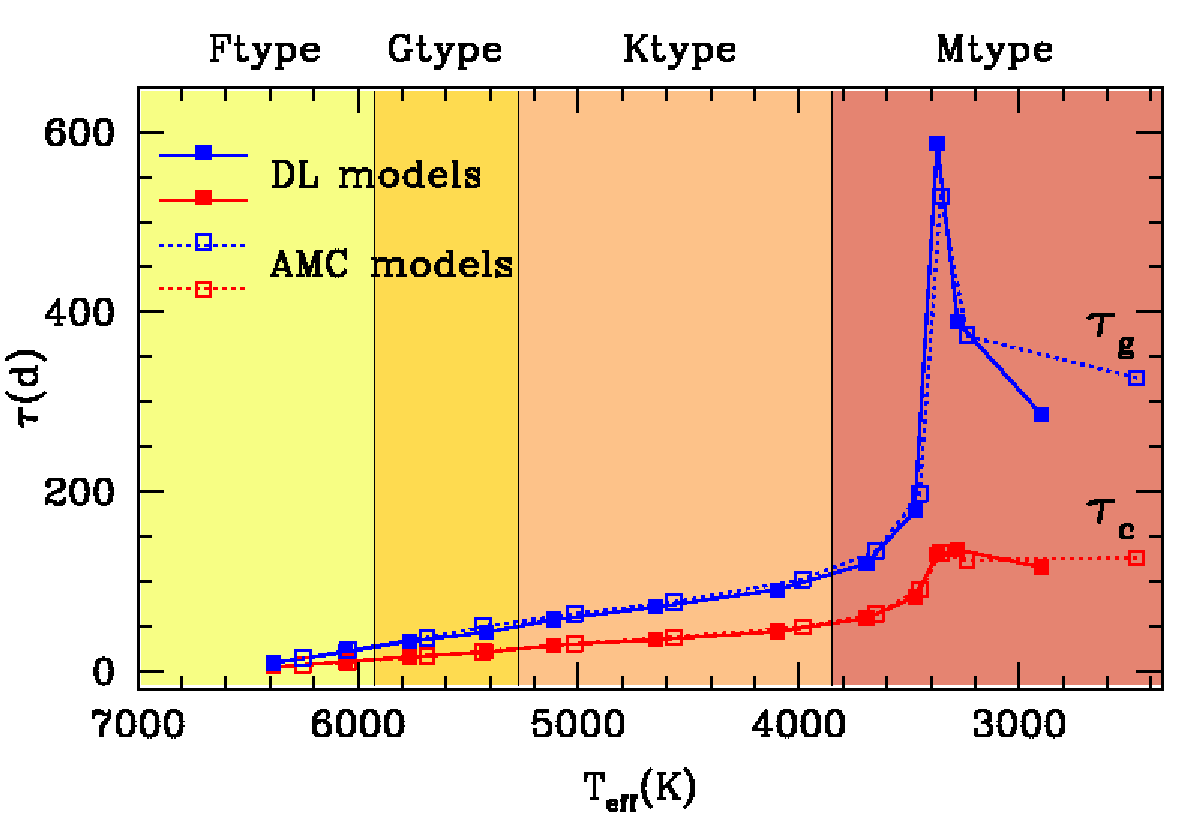}
\caption[Caption for LOF]{Similarly to Fig.\ 13 of \citet{irving23}, we show local (red) and global (blue) convective turnover times for AMC 
(dotted lines with open squares) and DL (solid lines with full squares) models
as a function of effective temperature at the ZAMS. We used an updated version
of \citet{pecaut2013} effective temperature-spectral type 
relation,
\scriptsize
https://www.pas.rochester.edu/~emamajek/EEM\_dwarf\_UBVIJHK\_colors\_Teff.txt. 
\label{taucxteff}
}
}
\end{figure}

After investigating the local convective turnover time behaviour as a function 
of age obtained by models with different DL parameters, 
Fig.~\ref{taucxteff}
shows 
how $\tau_{\rm c}$ and $\tau_{\rm g}$ vary with the 
effective temperature at the ZAMS. This plot is of particular interest
due to its adequacy in obtaining the Rossby number using a quantity 
obtainable
through observational data. As both global and local 
convective turnover times practically do not change during the main sequence
(see Figs.~\ref{taugage} and \ref{taulage}), their ZAMS values are representative 
for their entire main sequence. The effective temperatures at the ZAMS,
the ZAMS ages themselves, the global and local convective turnover 
times are slightly model dependent, and the differences increase with 
the decreasing effective temperatures (i.e. in the lower-mass regime).  

\section{Comparison with observations} \label{applications}

\begin{figure*}
\centering{
\includegraphics[width=09.1cm]{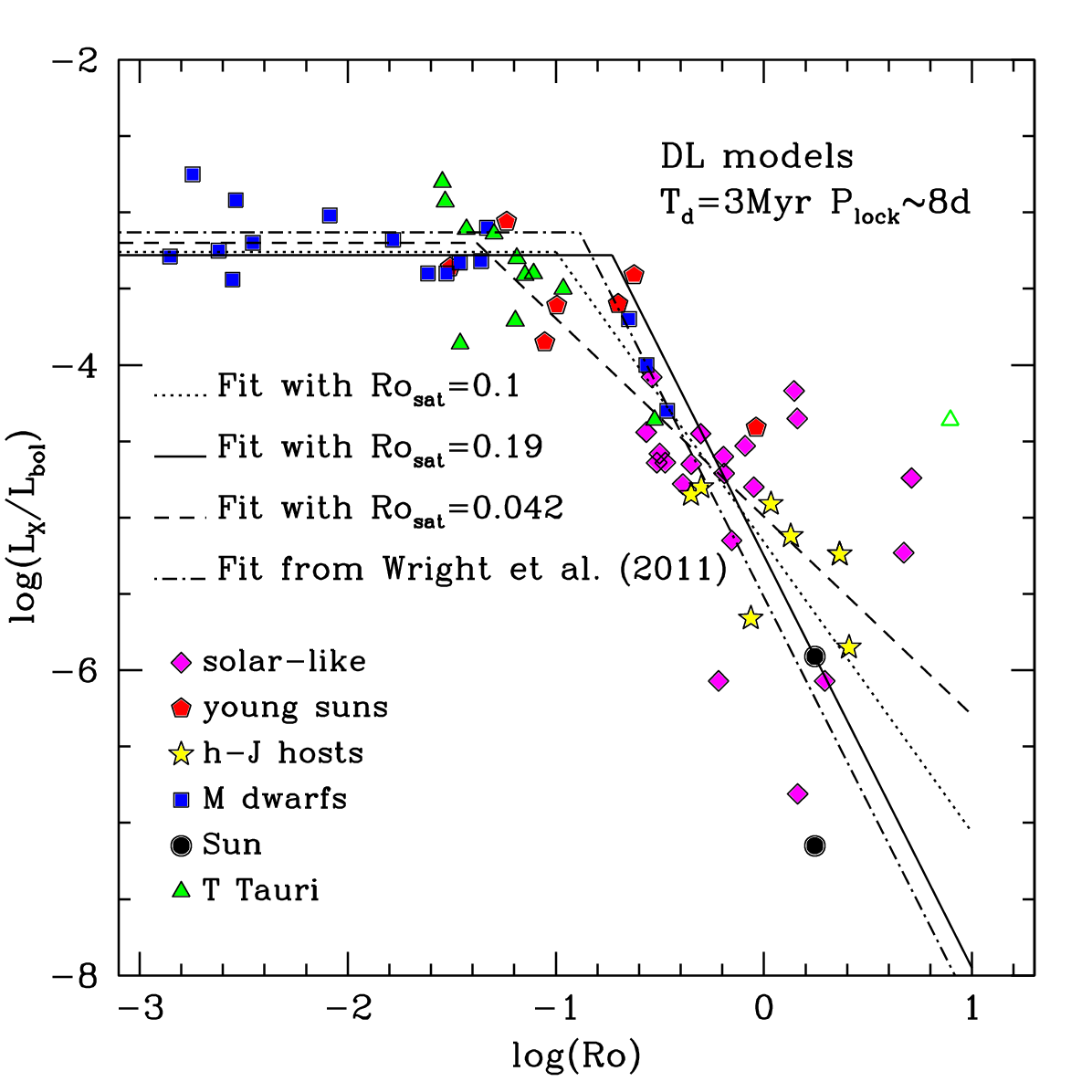}\hspace{-0.0860\textwidth}
\includegraphics[width=09.1cm]{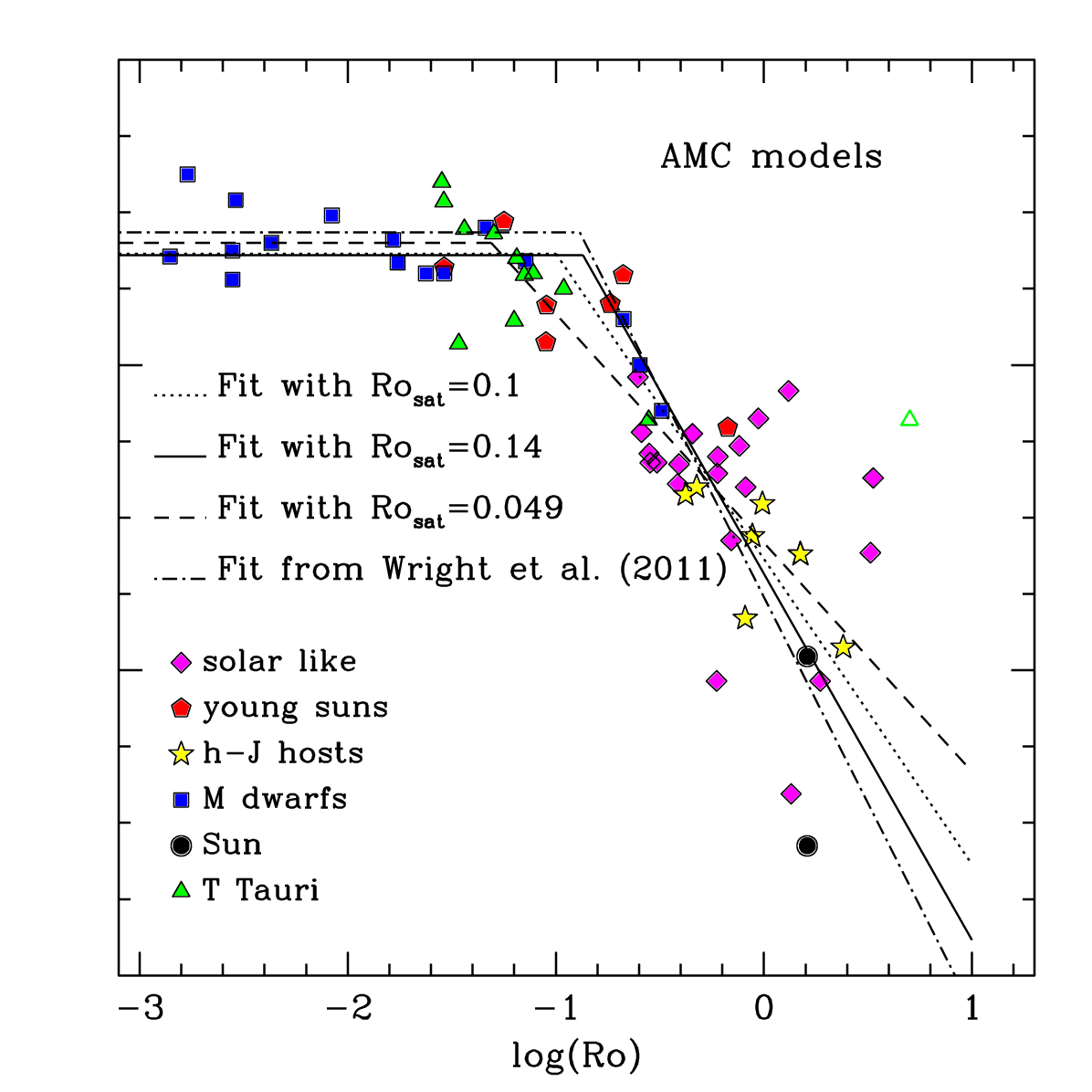}
\caption{$\log(L_{\rm X}/L_{\rm bol})$ $\times$ $\log(Ro)$
for stars in the sample of \citet{vidotto14}. In the left panel, $\tau_{\rm c}$ 
values, which enter in $Ro$ calculations, were obtained from DL models 
with $P_{\rm lock}\!\!\sim \!\!8$\,days and $T_{\rm disk}\!\!=\!\!3$\,Myr. In the right panel, $\tau_{\rm c}$ values
were obtained from AMC models.
Solar-like stars
are shown as \textcolor{magenta}{\rhombusfill} \hspace{-0.323cm}\rhombus, the Sun in its maximum and minimum levels of activity as \circletfill,
young suns as \textcolor{red}{\pentagofill}\hspace{-0.220cm}\pentago, hot-Jupiter hosts as \textcolor{yellow}{\starletfill}\hspace{-0.220cm}\starlet, M dwarfs as \textcolor{blue}{\squadfill}\hspace{-0.230cm}\squad\, and T\,Tauri
stars as \textcolor{green}{\trianglepafill}\hspace{-0.235cm}\trianglepa\,.
}
\label{lxlbolrodlkw}
}
\end{figure*}

\begin{figure*}
\centering{
\includegraphics[width=09.1cm]{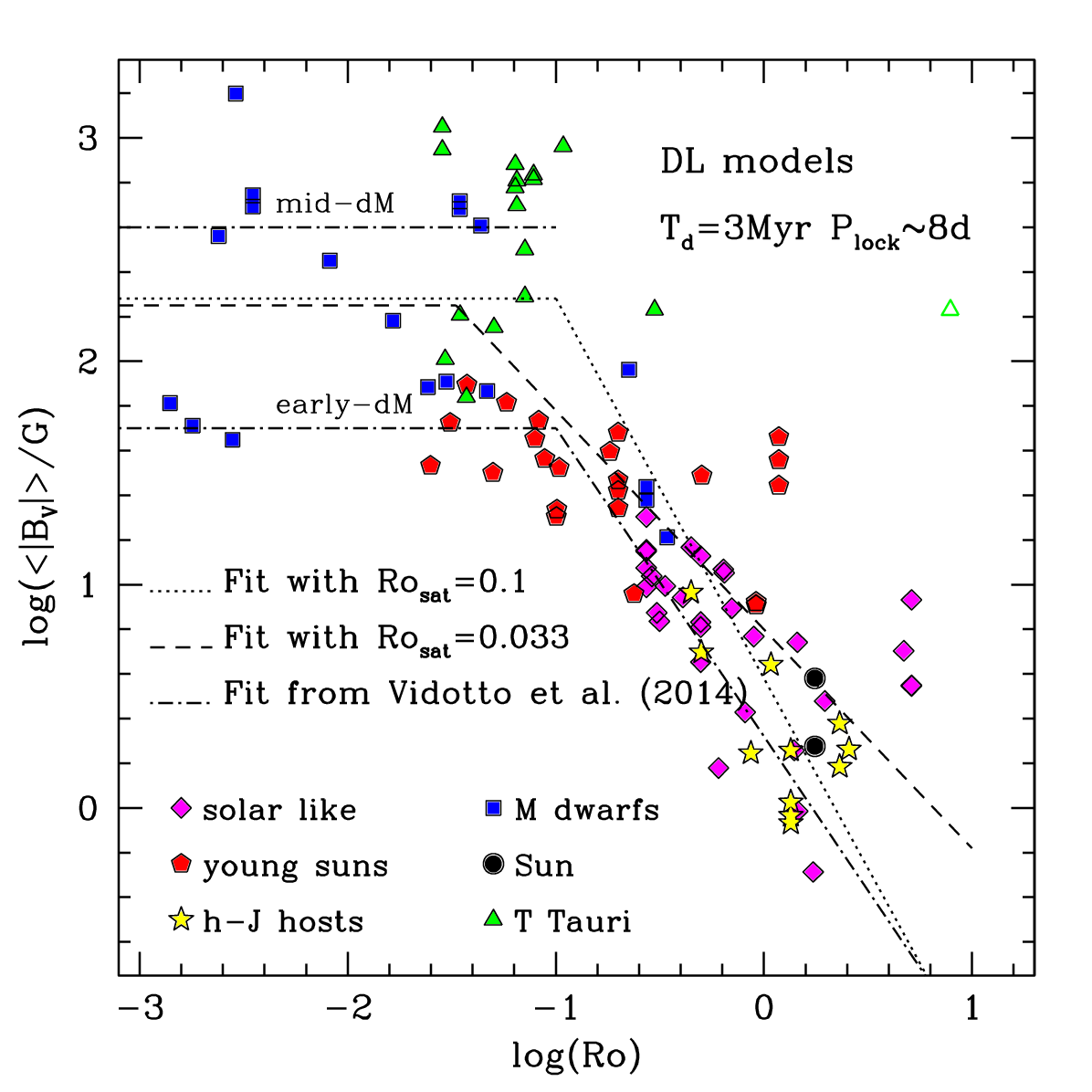}\hspace{-0.0860\textwidth}
\includegraphics[width=09.1cm]{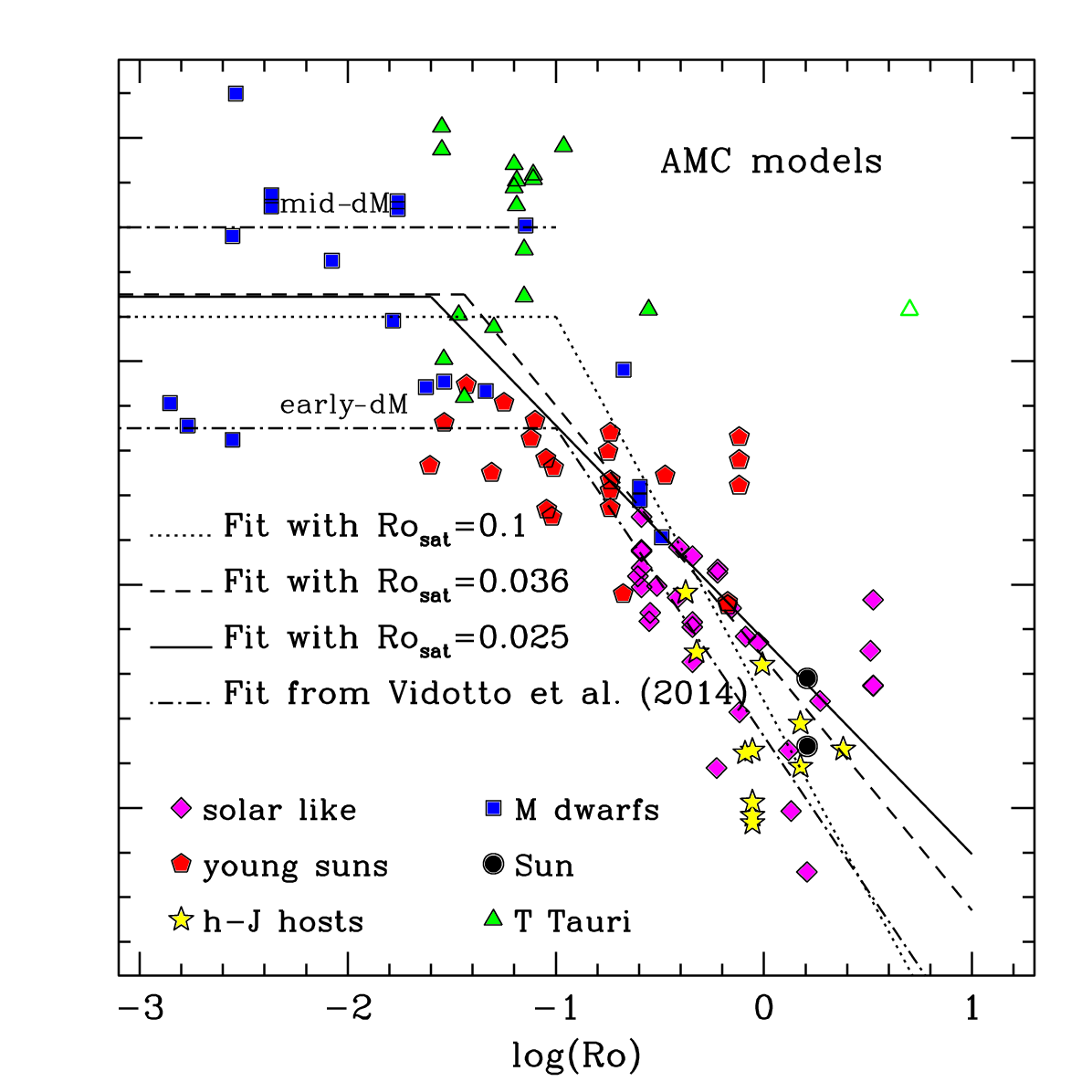}
\caption{$\log(\langle |B_{\rm V}|\rangle)$ $\times$ $\log(Ro)$
for stars in the sample of \citet{vidotto14}. 
In the left panel, the $\tau_{\rm c}$
values, which enter in $Ro$ calculations, were obtained from DL models
with $P_{\rm lock}\!\sim \!8$\,days and $T_{\rm disk}\!=\!3$\,Myr. In the right panel, $\tau_{\rm c}$ values
were obtained from AMC models. Symbols are
the same as in Fig.~\ref{lxlbolrodlkw}.}
\label{bvrodlkw}
}
\end{figure*}

In order to test our theoretical convective turnover times, we used them to calculate
$Ro$ for 73 late-F, G, K and M dwarf stars \citep[both pre-MS and MS from the sample of][]{vidotto14}
and to investigate the
magnetic activity-rotation relationship.
The sample is formed by solar-like stars [MS stars with
masses and ages in the ranges 0.66$\leq$$M/M_{\odot}$$\leq$1.34 and 
260$\leq$t(Myr)$\leq$8700, including the Sun itself], young suns 
[non-accreting pre-MS and young MS stars with masses and ages 
in the ranges 0.54$\leq$$M/M_{\odot}$$\leq$1.50 and 
10$\leq$t(Myr)$\leq$130],
hot-Jupiter (h$-$J) hosts [MS stars with 
masses and ages in the ranges 0.79$\leq$$M/M_{\odot}$$\leq$1.34 and 
600$\leq$t(Myr)$\leq$5000 that host planets as massive as Jupiter 
in very close orbits], M dwarfs [MS and non-accreting pre-MS 
stars with masses and ages in the ranges 0.10$\leq$$M/M_{\odot}$$\leq$0.75 
and 21$\leq$t(Myr)$\leq$1200] and Classical T\,Tauri stars [accreting 
pre-MS stars with masses and ages in the ranges 
0.65$\leq$$M/M_{\odot}$$\leq$2.00 and 1.4$\leq$t(Myr)$\leq$17]. 
\citet{vidotto14} separated their sample in these subsamples in order to 
investigate whether the stars in these subgroups showed any specific behaviour
in the rotation-magnetic activity diagram. 
They did not find any particular trend for any of these subgroups 
other than those already known: fully convective
stars tend to occupy the saturated region and partially convective ones 
occupy both regions.
Values of mass, age (except for 12 M dwarfs and 1 
T\,Tauri star), $P_{\rm rot}$, 
$L_{\rm X}/L_{\rm bol}$ and $\langle |B_{\rm V}|\rangle$ were taken from \citet{vidotto14}. The ages of the 12 M dwarfs
without age estimates in \citet{vidotto14} were determined using our evolution models. The age of 
the T\,Tauri star CV\,Cha used by \citet{vidotto14} is 4.8\,Myr,
taken from the value of 
(5$\pm$1)\,Myr
determined
by \citet{hussain09}. By using this age and our local convective turnover time, 
the Rossby number of CV\,Cha is
too
high, diverging considerably from the typical values of 
other T\,Tauri stars of our sample (as can be seen as open green triangles in Figs.~\ref{lxlbolrodlkw} 
and \ref{bvrodlkw}). We, then, took the inferior limit of the value published by \citet{hussain09}, 
i.e. 4\,Myr, as the age of CV\,Cha, which coincides, within the errors, with the value of (4.2$\pm$0.3)~Myr found using 
the {\ttfamily ATON} code. Next, given the stellar mass and age of each object, 
their Rossby numbers were obtained using the observed rotation periods from \citet{vidotto14} 
and the local convective turnover times produced by our DL and AMC models.  
The rotation-activity relation of our sample was analysed using two different
indicators of magnetic activity: $L_{\rm X}/L_{\rm bol}$ and $\langle |B_{\rm V}|\rangle$.
Before getting into the details of our calculations, we outline the general procedure
used for both indicators:
\vspace{-0.5\baselineskip}
\begin{enumerate}
  \item We initially fit the distribution of stars in the rotation-activity diagram with the
        following two-part, power-law functions,
\vspace{-0.005\textheight}        
        \begin{equation}
           \dfrac{L_{\rm X}}{L_{\rm bol}}  =
           \begin{cases}
              C\,Ro^{\beta},  & {\rm \text{ if }} Ro > Ro_{\rm sat}, \\
              \left(\dfrac{L_{\rm X}}{L_{\rm bol}}\right)_{\rm sat}, & {\rm \text{ if }} Ro\leq Ro_{\rm sat},\\
           \end{cases}
           \label{eqtwopart}
        \end{equation}
\vspace{-0.0065\textheight}        
        \begin{equation}
           \langle |B_{\rm V}|\rangle =
           \begin{cases}
              C\,Ro^{\beta},  & {\rm \text{ if }} Ro > Ro_{\rm sat}, \\
              \langle |B_{\rm V}|\rangle_{\rm sat}, & {\rm \text{ if }} Ro\leq Ro_{\rm sat},\\
           \end{cases}
           \label{eqtwopartbv}
        \end{equation}\noindent
        where $\beta$ is the power-law slope in a log-log plot for the unsaturated region and $C$ is a constant.
        Given an initial guess value for $Ro_{\rm sat}$, we determined either $(L_{\rm X}/L_{\rm bol})_{\rm sat}$
        or $\langle |B_{\rm V}|\rangle_{\rm sat}$, which are respectively the average values of $L_{\rm X}/L_{\rm bol}$
        and $\langle |B_{\rm V}|\rangle$ for $Ro$$\leq$$Ro_{\rm sat}$. For $Ro$$>$$Ro_{\rm sat}$,
        we fit the data by using an iterative linear regression fit in a log-log scale, keeping the constant
        coefficient fixed, so that the values of either $L_{\rm X}/L_{\rm bol}$ or $\langle |B_{\rm V}|\rangle$
        at $Ro_{\rm sat}$ are respectively equal to $(L_{\rm X}/L_{\rm bol})_{\rm sat}$ and
        $\langle |B_{\rm V}|\rangle_{\rm sat}$.
  \item Next we follow \citet{vidotto14} and \citet{jackson10} and fixed the saturation Rossby number 
        at its canonical value, $Ro_{\rm sat}$=0.1, first estimated by \citet{pizzolato03}, and determined the best
        parameters that fit the data to Eqs.\ \ref{eqtwopart} or \ref{eqtwopartbv}, i.e.\, $\beta$ and
        $(L_{\rm X}/L_{\rm bol})_{\rm sat}$ or $\langle |B_{\rm V}|\rangle_{\rm sat}$. This
        approach will be referred to as ``Method A''.      
  \item A second estimate of $Ro_{\rm sat}$ is obtained by using an iterative least squares method to find the value
        at which the standard deviation of the data reaches its minimum and, again, we get the corresponding
        values of $(L_{\rm X}/L_{\rm bol})_{\rm sat}$, $\langle |B_{\rm V}|\rangle_{\rm sat}$ and the best fitting
        slopes for the unsaturated region for both $L_{\rm X}/L_{\rm bol}$ and $\langle |B_{\rm V}|\rangle$ indicators.
        This approach will be referred to as ``Method B''.
  \item A third and last estimate for $Ro_{\rm sat}$ is made by choosing the value for which a least squares
        fit to the data results in the best fit to the Sun at its maximum, average and/or minimum activity levels. 
        Then, as previously, we obtain the corresponding values of
        $(L_{\rm X}/L_{\rm bol})_{\rm sat}$, $\langle |B_{\rm V}|\rangle_{\rm sat}$ and $\beta$. 
        This approach will be referred to as ``Method C''.
\end{enumerate}

\subsection{Using $L_{\rm X}/L_{\rm bol}$ as the magnetic activity indicator}\label{lxlbolind}

Among the 73 stars in the sample of \citet{vidotto14}, 11 have no fractional X-ray luminosity 
determination, reducing our sample to 62 stars in the analysis involving $L_{\rm X}/L_{\rm bol}$ 
as the magnetic activity indicator.
We initially analysed the rotation-activity relationship of our sample by using 
$L_{\rm X}/L_{\rm bol}$ and $Ro$ calculated with
$\tau_{\rm c}$ from DL models with $P_{\rm lock}\!\!\sim \!\!8$\,days and $T_{\rm disk}\!\!=\!\!3$\,Myr.
The left panel of Fig.~\ref{lxlbolrodlkw} shows our star sample in the 
$\log(L_{\rm X}/L_{\rm bol})$$\times$$\log(Ro)$ plane, with different symbols and colours.

For $Ro_{\rm sat}$$=$$0.1$ (Method A) the best fitting slope for the unsaturated part of the relation
(see Eq.~{\ref{eqtwopart}})
was found to be
$\beta\hphantom{,}$=$\hphantom{,}-$1.9$\pm$0.2, with the Pearson correlation coefficient 
$|\rho|$$=$0.66 and the saturation level was estimated to be 
$\log (L_{\rm X}/L_{\rm bol})_{\rm sat}$$=$$-$3.26$\pm$0.06 (dotted curve in the left panel 
of Fig.~\ref{lxlbolrodlkw}).
These values are consistent with those found by, e.g. \citet{wright11,wright18}.
For $Ro_{\rm sat}$ obtained through Method B,
we found $Ro_{\rm sat}$$=$0.042$\pm$0.003,
$\beta$$=$$-$1.3$\pm$0.2 (with $|\rho|$$=$0.80) and
$\log (L_{\rm X}/L_{\rm bol})_{\rm sat}$$=$$-$3.20$\pm$0.07 (dashed curve in the left panel
of Fig.~\ref{lxlbolrodlkw}). Although $\log (L_{\rm X}/L_{\rm bol})_{\rm sat}$ obtained in the 
last case is consistent with those found in the literature, 
the values of $\beta$ and $Ro_{\rm sat}$ are considerably higher and smaller, respectively. 
Notice that none of these methods fits the minimum or the maximum Sun (black circles).
By applying Method C, we could not fit the minimum Sun, while the Sun's 
position in its average level of activity (the average Sun) was only achieved with fitting 
parameters in disagreement with the literature. However, we could fit the maximum Sun and found
$Ro_{\rm sat}$$=$0.19, $\beta$$=$$-$2.7$\pm$0.4 (with $|\rho|$$=$0.61) and 
$\log (L_{\rm X}/L_{\rm bol})_{\rm sat}$$=$$-$3.28$\pm$0.05 (solid black line in the 
left panel of Fig.~\ref{lxlbolrodlkw}).

These values are consistent with those found in the literature, with the
value of $\beta$ matching that of
\citet{wright18} within the errors and 
coinciding with that of \citet{wright11}.
The value of $Ro_{\rm sat}$ is slightly larger than those of 
\citet{wright11,wright18} and $\log (L_{\rm X}/L_{\rm bol})_{\rm sat}$ 
is moderately smaller than (but still consistent with) those obtained by 
both mentioned works. 
For these reasons, and also because this fit reproduces the maximum Sun's position in the 
rotation-activity diagram with consistent parameters, we consider it the best fit using
DL models.
For comparison, the best fit found by \citet{wright11}, obtained using another sample of stars, is also shown in the left 
panel of Fig.~\ref{lxlbolrodlkw}.
Table \ref{wrighttab} shows the fit parameters found in this subsection 
(using $\tau_{\rm c}$ provided by DL models) and those found by \citet{wright11} 
and \citet{wright18}.

\begin{table}
\caption{Fit parameters of the rotation-magnetic activity relationship
in this work, W11 \citep{wright11}, W18 \citep{wright18} and G23 \citep{galvao23}.}
\label{wrighttab}
\centering
{\footnotesize
\advance\tabcolsep by -4.0pt
\begin{tabular}{lrrr}
\hline \hline
\\ [-08pt]
~~~~~~~~~~~~Work       & $Ro_{\rm sat}$ ~~~~~~  &  $\beta$ ~~~~~~      & $\!\log\biggl( \frac{L}{L_{\rm bol}} \biggr)_{\rm sat}$ \\ [+6.0pt] \hline
\\ [-08pt]
DL mod., Meth.\ A    &    0.1                 & $-1.9$$\pm$$0.2$          &$-3.26$$\pm$$0.06$                                       \\ [+1.0pt]
\hline
\\ [-08pt]
DL mod., Meth.\ B    & $0.042$$\pm$$0.003$       & $-1.3$$\pm$$0.2$          &$-3.20$$\pm$$0.07$                                       \\ [+1.0pt]
\hline
\\ [-08pt]
DL mod., Meth.\ C    &    0.19                & $-2.7$$\pm$$0.4$          &$-3.28$$\pm$$0.05$                                       \\ [+1.0pt]
\hline
\\ [-08pt]
AMC mod., Meth.\ A   &    0.1                 & $-2.0$$\pm$$0.2$          &$-3.27$$\pm$$0.06$                                       \\ [+1.0pt]
\hline
\\ [-08pt]
AMC mod.\, Meth.\ B~~~~~~   &$0.049$$\pm$$0.006$       & $-1.5$$\pm$$0.2$          &$-3.20$$\pm$$0.07$                                       \\ [+1.0pt]
\hline
\\ [-08pt]
AMC mod., Meth.\ C   &    0.14                & $-2.4$$\pm$$0.3$          &$-3.28$$\pm$$0.05$                                       \\ [+1.0pt]
\hline
\\ [-08pt]
W11                    & $0.13$$\pm$$0.02$         & $-2.7$$\pm$$0.13$         &$-3.13$$\pm$$0.08$                                       \\[+1.0pt] 
\hline
\\ [-09pt]
W18                    & $0.14^{+0.02}_{-0.04}$ & $-2.3^{+0.4}_{-0.6}$   &$-3.05^{+0.05}_{-0.06}$                               \\[+1.0pt] 
\hline
\\ [-08pt]
G23                    & $0.045$$\pm$$0.001$       & $-1.30$$\pm$$0.08$        &$-3.114$$\pm$$0.006$                               \\[+1.0pt]  \hline
\end{tabular}
}
\end{table}

Using values of $\tau_{\rm c}$ calculated with AMC models, we obtained
the Rossby number of all stars in our sample and displayed them in the rotation-activity diagram,
as shown in the right panel of Fig.~\ref{lxlbolrodlkw}. 
As for the DL case, we fit the AMC data
with three values of $Ro_{\rm sat}$. For $Ro_{\rm sat}$$=$$0.1$ (Method A) we found a power index $\beta$$=$$-2.0$$\pm$$0.2$ (with $\rho$$=$$0.69$) and
a saturation level of $\log(L_{\rm X}/L_{\rm bol})_{\rm sat}$$=$$-3.27$$\pm$$0.06$ (dotted curve 
in the right panel of Fig.~\ref{lxlbolrodlkw}). These values are consistent with
those found in the literature, but $\log(L_{\rm X}/L_{\rm bol})_{\rm sat}$ is slightly low. 
Then, by using Method B, we found $Ro_{\rm sat}$$=$0.049$\pm$0.006, 
$\beta$$=$$-$1.5$\pm$0.2 (with $|\rho|$$=$$0.81$) and $\log(L_{\rm X}/L_{\rm bol})_{\rm sat}$$=$$-3.20$$\pm$$0.07$  
(dashed curve in the right panel of Fig.~\ref{lxlbolrodlkw}). These values of 
$Ro_{\rm sat}$ and $\beta$ are compatible with those found in the literature,
despite being marginally lower and higher, respectively, while the saturated value of 
$L_{\rm X}/L_{\rm bol}$ matches the one found by \citet{wright11} within the 
uncertainties. Nearly identical parameters were
obtained for this sample by \citet{galvao23}, who used $\tau_{\rm c}$ determined with evolutionary tracks from the {\ttfamily ATON} code
(similar to AMC models) and a Markov Chain Monte Carlo (MCMC) fitting method.
Again, as can be seen from the right panel of Fig.~\ref{lxlbolrodlkw}, 
the dotted and dashed curves do not fit the Sun. 
Using Method C, we obtained a saturation Rossby number of 
$Ro_{\rm sat}$$=$0.14. This method fits the maximum Sun, provides $\log(L_{\rm X}/L_{\rm bol})_{\rm sat}$$=$$-$3.28$\pm$0.05 
and $\beta$$=$$-$2.4$\pm$0.3 (with $|\rho|=0.66$) and is shown as a solid curve in the right panel of 
Fig.~\ref{lxlbolrodlkw}. 
The value of $Ro_{\rm sat}$ agrees with those found by \citet{wright11} 
and \citet{wright18} and is moderately larger than that
found by \citet{landin23}, while the value of $\beta$ agrees, within the 
uncertainties, with those found by \citet{wright11}, \citet{wright18} and 
\citet{landin23}, that used much larger samples of stars (824 stars for the 
former work and 847 stars for the latter ones). 
Likewise, this value of $\log(L_{\rm X}/L_{\rm bol})_{\rm sat}$ is consistent with 
those found by these authors, although a little smaller.
As in the case of DL models, this method cannot fit the minimum Sun 
and fits the average Sun with conflicting parameters. For the same reasons
discussed in the analysis of the rotation-activity diagram with $\tau_{\rm c}$
obtained by DL models, the parameters that fit the maximum Sun seem to be the more 
reliable ones when AMC models are used to calculate $\tau_{\rm c}$ and they are shown in Table~\ref{wrighttab}, together with the other AMC fits obtained in this subsection.
For comparison purposes, the best fit found by \citet{wright11} using another sample of stars is also shown in the right panel of 
Fig.~\ref{lxlbolrodlkw}. 

The values obtained for $\beta$ and $\log(L_{\rm X}/L_{\rm bol})_{\rm sat}$ with 
$\tau_{\rm c}$ calculated by AMC models agree, within the errors, with 
the corresponding values obtained by DL models. 
Regarding
the fits obtained with Methods A and B, the
$\beta$ 
values
are slightly smaller for AMC models and the opposite situation is 
observed by the fits which reproduce the Sun's position in its maximum activity level (Method C).
For both sets of models, we consider that the fits which better predict the
solar magnetic activity level are the best ones. Besides 
fitting the Sun, they present $Ro_{\rm sat}$ and $\beta$ in accordance with
what was found in the works of \citet{wright11,wright18}.

As we can see in Fig.~\ref{taulage}, $\tau_{\rm c}$ produced by AMC models 
are higher than those generated by DL models for most stellar ages, implying in 
smaller Rossby numbers. In fact, the average value of Rossby numbers obtained 
with DL models, $\langle Ro_{\rm DL}\rangle$, is 21\% higher than that obtained 
with AMC models, $\langle Ro_{\rm AMC}\rangle$, which 
causes a slight global shift of the star distribution to the left 
in the rotation-activity diagram.
However, this does not translate into 
a lower saturation Rossby number in our least squares fit,  
due to the high dispersion of the data. 
The fact that $\langle Ro_{\rm DL}\rangle$ is larger than 
$\langle Ro_{\rm AMC}\rangle$ indicates that stars that had experienced 
(or are experiencing) a locking phase present higher Rossby numbers and then, 
lower dynamo numbers, which means lower efficiency in the dynamo generation
process. 

It is already known that internal redistribution of angular momentum 
and angular momentum loss by stellar magnetised winds can significantly
affect the stellar angular momentum evolution. As for the moment, we cannot 
consider yet these effects in our results because the 
{\ttfamily ATON} code version that takes the disk-locking mechanism into 
account is currently implemented only for rotational scheme 1 (solid body rotation), 
while the effects related to angular momentum variation are implemented only for 
rotational scheme 3 (differential rotation in radiative regions and
solid body rotation in convective zones).
At 100\,Myr, $\tau_{\rm c}$ obtained for 0.4 and 1.0\,M$_{\odot}$ 
models considering redistribution of 
angular momentum and angular momentum loss by winds is around 94\% of the 
value obtained when considering solid body rotation with 
no effects related to angular momentum variability. As far as we could foresee, 
our DL and AMC models would be affected roughly the same way by the 
inclusion of such effects, keeping invariable the relative differences 
between our results obtained with the two sets of models. Including 
these effects would cause DL and AMC models to produce $\tau_{\rm c}$ 
values a few percent lower. Consequently, the Rossby numbers obtained 
for the stars of our sample would be slightly higher (shifting the
star distributions in both panels of Fig.~\ref{lxlbolrodlkw} to the right), 
which would probably slightly increase $Ro_{\rm sat}$.
Nevertheless, we plan to improve the {\ttfamily ATON} code in order to fully account
for these effects.

\subsection{Using $\langle |B_{\rm V}|\rangle
$ as the magnetic activity indicator}\label{bvind}

The sample of \citet{vidotto14} consists of 104 magnetic maps of 73 stars, constructed with 
observations made at multiple epochs, resulting in 102 pairs of measurements of unsigned 
average large-scale surface magnetic fields and rotation periods, which allow us to use
$\langle |B_{\rm V}|\rangle$ as the magnetic activity indicator in the analysis of the
rotation-activity relation of our sample.
Initially, we used Rossby numbers 
calculated by DL models with $P_{\rm lock}$$\sim$8\,days and $T_{\rm disk}$=3\,Myr. 
In the left panel of Fig.~\ref{bvrodlkw}, we show the stars of our sample in the 
$\log(\langle |B_{\rm V}|\rangle)$ $\times$ $\log(Ro)$ plane with the same symbols 
and colours used in Fig.~\ref{lxlbolrodlkw}.

The same general procedure used for the $\log(L_{\rm X}/L_{\rm bol})$ indicator was
adopted and, for $Ro_{\rm sat}$$=$$0.1$ (Method A),
we found $\beta$$=$$-1.70$$\pm$$0.15$, 
as the best fitting slope for the unsaturated part of the relation (see Eq.\ \ref{eqtwopartbv}), with 
$\rho$$=$$0.66$, and the saturation level of activity was found to be 
$\log(\langle |B_{\rm V}|\rangle /{\rm G})_{\rm sat}$=$2.28$$\pm$$0.06$ (dotted curve in the 
left panel of Fig.~\ref{bvrodlkw}). The value of $\beta$ is nearly in agreement
with that of \citet{vidotto14} within the uncertainties. Our value of 
$\log(\langle |B_{\rm V}|\rangle)_{\rm sat}$ is between the saturation value found by 
\citet{vidotto14} for the early M dwarfs (early-dM, $M$$\ge$$0.4\,$M$_{\odot}$, 
$\log(\langle |B_{\rm V}|\rangle /{\rm G})_{\rm sat}$$=$$1.7$) and the value 
found for mid M dwarfs (mid-dM, 0.2$<$$M/M_{\odot}$$<$0.4, 
$\log(\langle |B_{\rm V}|\rangle /{\rm G})_{\rm sat}$$=$$2.6$).
These two saturation levels are shown as horizontal
dot-dashed lines in both panels of Fig.~\ref{bvrodlkw}.
With Method B, we found $Ro_{\rm sat}$$=$$0.033$$\pm$$0.012$, $\beta$$=$$-0.98$$\pm$$0.11$ (with 
$|\rho$$=$$0.81|$) and $\log(\langle |B_{\rm V}|\rangle /{\rm G})_{\rm sat}$$=$$2.25$$\pm$$0.14$
(dashed curve in the left panel of fig.~\ref{bvrodlkw}). These values of 
$\beta$ and $\langle |B_{\rm V}|\rangle_{\rm sat}$ are consistent
with those found by \citet{vidotto14}, but our $\beta$ is slightly higher and 
our $Ro_{\rm sat}$ is considerably lower. According to \citet{see19}, who
also found a $Ro_{\rm sat}$ smaller than \citet{vidotto14} but a little
larger than this work, obtaining a smaller value of $Ro_{\rm sat}$ could be due
to a number of reasons. They mentioned that a poorly constrained level of the
saturation field strength and differences in the way to calculate the convective
turnover times can affect the $Ro_{\rm sat}$ value. In addition, they 
emphasise that different activity indicators could saturate at different 
Rossby numbers.
As can be seen in the left panel of 
Fig.~\ref{bvrodlkw}, our fit obtained through Method B fits the Sun in its maximum level of activity, 
while our fit with $Ro_{\rm sat}$$=$$0.1$ (Method A) nearly fits the Sun in its minimum.
Although Method B fitted the maximum Sun, we applied Method C and, with it, we were also able to 
reproduce the minimum and the average Sun, in addition to the maximum Sun, with plausible parameters, similar to those
found by \citet{see19} and \citet{galvao23}. We decided to present only the parameters that fit 
the maximum Sun because this was the only level of solar activity that we were able to reproduce with 
Method C, employing the two magnetic activity indicators used in this work.
For comparison, the best fit found by \citet{vidotto14} is also shown in the left 
panel of Fig.~\ref{bvrodlkw} as dot-dashed lines. Here, it is worth mentioning that our fit includes
all T\,Tauri stars and the 12 M dwarfs without age estimates in \citet{vidotto14}
but these stars did not enter in their fit.
Table \ref{vidseetab} shows the fit parameters found in this subsection 
(using $\tau_{\rm c}$ provided by DL models) and those found by \citet{vidotto14} and 
\citet{see19}.

\begin{table}
\caption{Fit parameters of the rotation-magnetic activity relation
in this work, \citet{vidotto14}, \citet{see19} and \citet{galvao23}.}
\label{vidseetab}
\centering
{\footnotesize
\advance\tabcolsep by -4.0pt
\begin{tabular}{lrrr}
\hline \hline
\\ [-08pt]
~~~~~~~~Work            & $Ro_{\rm sat}$~~~~~~   &  $\beta$~~~~~~~~     & $\log\biggl(\frac{\langle |B_{\rm V}| \rangle}{G}\biggr)_{\rm sat}$ \\ [+6.0pt] \hline 
\\ [-08pt]
DL mod., Meth.\ A     &    0.1                 & $-1.70\pm 0.15$        & $2.28\pm 0.06$                                       \\  [+1.0pt]
\hline
\\ [-08pt]
DL mod., Meth.\ B     & $0.033\pm 0.012$       & $-0.98\pm 0.11$        & $2.25\pm 0.14$                                       \\  [+1.0pt]
\hline
\\ [-08pt]
AMC mod., Meth.\ A    &    0.1                 & $-1.72\pm 0.15$        & $2.20\pm 0.09$                                       \\  [+1.0pt]
\hline
\\ [-08pt]
AMC mod., Meth.\ B    & ~~$0.036\pm 0.008$       & $-1.13\pm 0.11$        & $2.30\pm 0.12$                                       \\  [+1.0pt]
\hline
\\ [-08pt]
AMC mod., Meth.\ C    &    0.25                & $-0.96\pm 0.10$        & $2.29\pm 0.15$                                       \\  [+1.0pt]
\hline
\\ [-08pt]
Vidotto et al. (2014)   &   0.1                  & $-1.38\pm 0.14$        & 1.7 (early-dM)                                       \\ [-1.0pt]
                &                                &                        &  2.6 (mid-dM)                                       \\ [+1.0pt] \hline
\\ [-08pt]
See et al.\ (2019)      & $0.06\pm 0.01$         & $-1.40\pm 0.10$        & $2.41\pm 0.12$                               \\ [+1.0pt]
\hline
\\ [-08pt]
Galv\~ao et al.\ (2023) & $0.059\pm 0.002$       & $-1.23\pm 0.01$        & $2.19\pm 0.01$                               \\ [+1.0pt]
\hline
\end{tabular}
}
\end{table}

Finally,
we
investigated the rotation-activity relationship of our sample by using
our local convective turnover times obtained with AMC models
to calculate the Rossby number of all stars. Our sample is displayed in the
rotation-activity diagram shown in the right panel of Fig.~\ref{bvrodlkw} and the
data were fit with the two-part power-law function expressed in Eq.~\ref{eqtwopartbv}.
We again performed three fits to our data by imposing
$\langle|B_{\rm V}|\rangle$$=$$\langle|B_{\rm V}|\rangle_{\rm sat}$ at $Ro_{\rm sat}$.
For $Ro_{\rm sat}$$=$$0.1$ (Method A), we found $\beta$$=$$-1.72$$\pm$$0.15$
(with $\rho$$=$$0.68$) and $\log(\langle|B_{\rm V}|\rangle /{\rm G})_{\rm sat}$$=$$2.20$$\pm$$0.09$
(dotted curve in the right panel of Fig.~\ref{bvrodlkw}). This value of
$\langle|B_{\rm V}|\rangle_{\rm sat}$ is between the values proposed by
\citet{vidotto14} and in agreement with that of \citet{see19} within the 
uncertainties and this value of $\beta$ is consistent with that published by
both works.
Next, we applied Method B and found $Ro_{\rm sat}$$=$$0.036\pm 0.08$,
$\beta$$=$$-1.13$$\pm$$0.11$ (with $\rho$$=$$0.80$) and
$\log(\langle|B_{\rm V}|\rangle /{\rm G})_{\rm sat}$$=$$2.30$$\pm$$0.12$ (dashed curve in the right
panel of Fig.~\ref{bvrodlkw}). Although this $Ro_{\rm sat}$ is considerably smaller
than others found in the literature 
\citep[except that of][]{see19}, this value of $\beta$ matches that of \citet{vidotto14}
within the errors and nearly matches that of \citet{see19}, and the value of 
$\langle|B_{\rm V}|\rangle_{\rm sat}$ is consistent with
those found by
both works.
Even by using a different
fitting method (MCMC), \citet{galvao23} also
found similar parameters as in our analysis. As one can see in the right panel of Fig.~\ref{bvrodlkw},
the dotted curve nearly fits the minimum Sun and the dashed curve is close to the
maximum Sun. So, our last fit aimed to get a better fit to the maximum Sun 
by means of Method C, resulting in
$Ro_{\rm sat}$$=$$0.025$, which is considerably lower than the
values found in the literature. This last fit also provides $\beta\!\!=\!\!-0.96\pm0.10$
and $\log(\langle|B_{\rm V}|\rangle /{\rm G})_{\rm sat}\!\!=\!\!2.29\pm0.15$ (with $|\rho|$=0.81, seen
in solid line in the right panel
of Fig.~\ref{bvrodlkw}). These values are consistent with those found by \citet{vidotto14} and \citet{see19},
although the value of $\beta$ is somewhat higher. 
With
Method C, we could also fit both the minimum and the average Sun with consistent parameters.
As in the previous analysis, the parameters that fit the Sun in the study 
of the rotation-activity relation using $\langle |B_{\rm V}| \rangle$ as an 
activity indicator and $\tau_{\rm c}$ generated by AMC models
(all shown in Table \ref{vidseetab}) are also 
considered the more reasonable ones.
For comparison, the best fit found
by \citet{vidotto14} is also shown in the right panel of Fig.~\ref{bvrodlkw} as dot-dashed
lines.

The values 
obtained for $\beta$ and $\langle|B_{\rm V}|\rangle_{\rm sat}$ with
$\tau_{\rm c}$ calculated by AMC models agree, within the errors, with
those obtained by the corresponding fits with $\tau_{\rm c}$ calculated by DL models.

As described in Subsection~\ref{lxlbolind}, the values of Rossby numbers obtained with
$\tau_{\rm c}$ produced by AMC models are lower than those
generated by DL models and this manifests itself as a
slight
shift of
the stars distribution to the left in the right panel of 
Fig.~\ref{bvrodlkw} relative to those in the same figure's left panel.
For this subsample, 
$\langle Ro_{\rm DL}\rangle$=1.27\,$\langle Ro_{\rm AMC}\rangle$, what
would again imply that stars which pass through a locking phase would have a smaller
dynamo number and, consequently would operate a less efficient dynamo. However,
as in Subsection~\ref{lxlbolind}, $Ro_{\rm sat,DL}$$<$$Ro_{\rm sat,AMC}$, 
probably due to the high dispersion of the data. Lastly, the inclusion of 
redistribution of angular momentum and angular momentum loss by winds would 
affect the results obtained with DL and AMC models in a similar fashion, 
slightly increasing the stars Rossby numbers and moving the whole star 
distributions in both panels of Fig.~\ref{bvrodlkw} 
to the right, as discussed in the previous subsection. 

\section{Conclusions} \label{conclusions}
We present new sets of pre-MS and MS evolutionary tracks considering the 
disk-locking mechanism and the conservation of angular momentum, including Rossby 
numbers, global and local convective turnover times in the mass range of 
0.1-1.2\,M$_{\odot}$.
We compared $\tau_{\rm g}$, $\tau_{\rm c}$ and $\varv_{\rm c}$ produced by the
two sets of models considering both evolutionary phases (pre-MS and MS). 
The differences are smaller in the early pre-MS ($\sim$1\%) and increase 
with stellar age, reaching 8-10\% for ages greater than 100\,Myr. The 
largest differences are found in stars with masses higher than or equal to 1\,${\rm M_{\odot}}$.

In the MS, models simulating the DL
mechanism produce $\varv_{\rm c}$ higher than models conserving angular momentum.
The opposite behaviour is observed for $\tau_{\rm g}$ and $\tau_{\rm c}$,
except for 0.2 and 0.3\,M$_{\odot}$ models. 
More specifically, our results point out that stars which experience a locking 
phase reach the
MS
with smaller values of $\tau_{\rm c}$, larger
Rossby numbers and smaller dynamo numbers, implying that their dynamo 
processes are less efficient in comparison
with those stars that evolve conserving angular momentum since the beginning.
By varying the initial angular momentum and the duration of the disk phase, we 
noticed that the higher the initial velocity and $T_{\rm disk}$, the shorter
$\tau_{\rm c}$.
As the local convective turnover time is used to obtain the Rossby number, 
this suggests that properties of the disk phase of the star affect its 
$Ro$ and its position in the rotation-activity diagram in the MS.

In the sequence, we used our theoretical convective turnover times to calculate
$Ro$ for 73 stars from the sample of \citet{vidotto14} to investigate the
magnetic activity-rotation relationship. We plot the data in 
two versions of the rotation-activity diagram 
($L_{\rm X}/L_{\rm bol}$$\times$$Ro$ and 
$\langle |B_{\rm V}|\rangle$$\times$$Ro$) 
which show
the typical saturated and unsaturated regimes.
We fit the data with two-part 
power-law functions (Eqs.\,\ref{eqtwopart} and \ref{eqtwopartbv}) using three
different methods. By considering the fit in the 
$L_{\rm X}/L_{\rm bol}$$\times$$Ro$ diagram (with $\tau_{\rm c}$ calculated by 
DL models) which 
reproduces the active Sun’s position (Method C), we found $Ro_{\rm sat}$=0.14,
$\log(L_{\rm X}/L_{\rm bol})_{\rm sat}$=$-3.28$$\pm$0.05 and the inclination of 
the unsaturated region was found to be $\beta$=$-$2.4$\pm$0.3. By considering 
the best fit in the $\langle |B_{\rm V}|\rangle$$\times$$Ro$ diagram 
which reproduces the Sun’s position 
in its maximum of activity (obtained with Method B), we found 
$Ro_{\rm sat}$=0.033$\pm$0.012, 
$\log(\langle |B_{\rm V}|\rangle /{\rm G})_{\rm sat}$= 2.25$\pm$0.15 and the 
power index of the unsaturated region was found to be $\beta$=$-$0.98$\pm$0.11.   These parameters are consistent to those usually found in the literature.
We caution that these results are based on a relatively small sample, specially
when it is compared to other similar works (for instance, \citealt{wright11,wright18} and \citealt{newton17}).
In addition, $\tau_{\rm c}$ in Subsections~\ref{lxlbolind} and \ref{bvind} were 
obtained assuming either that all stars had gone through a disk phase (with about 
the same locking period and during the same disk lifetime) or that they all
evolved conserving angular momentum from the beginning.
The most likely scenario is that some stars would have evolved with a very short disk-locking phase ($T_{\rm disk}$$\lesssim$$10^5$\,yr) and 
can be considered as having evolved without a circumstellar disk since the
start
of their 
evolutions, while others would have spent some time in a locking phase with
different locking periods and different disk lifetimes.

Finally, we speculated how internal redistribution of angular 
momentum and angular momentum loss by magnetised stellar winds would affect
our results. We took such effects into account in 0.4 and 1.0\,M$_{\odot}$ 
models which rotate according to rotational scheme 3 and conserve angular 
momentum during all evolutionary phases and showed that it 
reduces $\tau_{\rm c}$ in 6\% at 100\,Myr. We found no reason why these
effects would affect DL models differently than AMC models (remembering 
that both were modelled with rigid body rotation). Therefore, the general 
behaviour we would expect for the stars in our sample in 
Section~\ref{applications} is that they would have smaller $\tau_{\rm c}$, which 
would imply in a larger $Ro$ and in a less efficient dynamo ($N_{\rm D}\propto Ro^{-2}$).

\section*{Data Availability}
The complete version of Table \ref{tabtrack} is only available in electronic form at the CDS 
via anonymous ftp to cdsarc.u-strasbg.fr (130.79.128.5) or via 
http://cdsweb.u-strasbg.fr/cgi-bin/qcat?J/A+A/.

\begin{acknowledgements} 
The authors thank Drs.\ Francesca D'Antona (INAF-OAR, Italy) and Italo Mazzitelli
(in memoriam) for granting them full access to the {\ttfamily ATON} evolutionary
code.  
They are also grateful to an anonymous referee
for his/her many comments and suggestions that
helped to improve this work.
Financial support from
the Brazilian agencies CAPES, CNPq and FAPEMIG is gratefully acknowledged.
\end{acknowledgements}

%
%

\end{document}